\documentclass[twocolumn,secnumarabic,amssymb, amsmath, tightenlines,nobibnotes, aps, prl,longbibliography]{revtex4-1}

\usepackage[pdftex,colorlinks=true,pdfstartview=Fit]{hyperref}
\usepackage{amsmath,amsfonts}
\usepackage{graphicx}  
\usepackage{wasysym}
\usepackage{amsmath,amssymb,psfrag,textcomp}
\usepackage{amsthm,upgreek,bm}   
\usepackage{bbm}
\usepackage{color}			
\usepackage{pdfpages}

\newcommand{\bG}{\ensuremath{\mathbf{G}}}

\newcommand{\bone     }{\mbox{\boldmath$1$}}

\newcommand{\bA     }{\mbox{\boldmath$A$}}

\usepackage{color}

\begin{document}
\title{Eigenvalue Outliers of  non-Hermitian Random Matrices with a Local Tree Structure}
\author{Izaak Neri$^{1,2}$ and Fernando Lucas Metz$^{3}$} 
\affiliation{ 
${^1}$Max Planck Institute for the Physics of Complex Systems, N{\"o}thnitzerstra{\ss}e 38, 01187 Dresden, Germany.\\
${^2}$ Max Planck Institute of Molecular Cell Biology and Genetics, Pfotenhauerstra{\ss}e 108, 01307 Dresden, Germany.\\
$^{3}$ Departamento de F\'isica, Universidade Federal de Santa Maria, 97105-900 Santa Maria, Brazil.
}
\begin{abstract}  
Spectra of sparse non-Hermitian random matrices determine the dynamics of complex processes on graphs. Eigenvalue outliers in the spectrum are of particular interest, since they determine the stationary state and the stability of dynamical processes. We present a general and exact theory for the eigenvalue outliers of random matrices with a local tree structure. For adjacency and Laplacian matrices of oriented random graphs, we derive analytical expressions for the eigenvalue outliers, the first moments of the distribution of eigenvector elements associated with an outlier, the support of the spectral density, and the spectral gap. We show that these spectral observables obey universal expressions, which hold for a broad class of oriented random matrices.
\end{abstract} 
\pacs{02.50.-r, 02.10.Yn, 89.75.Hc}
\maketitle     

\paragraph{Introduction}
Directed graphs   represent  the  causal relations between  the degrees of freedom of   a dynamical system.  Neural networks, transportation networks, and the Internet are examples of systems modelled by directed graphs. 
The dynamics of processes governed through directed graphs can be modeled with sparse non-Hermitian matrices, for example, Markov matrices define the dynamics of stochastic processes ~\cite{seneta2006non, levin2009markov}, and Jacobian matrices determine the stability of dynamical systems ~\cite{seydel2009practical}.

The dynamics of complex systems can be studied from the spectra of   sparse non-Hermitian  {\it random} matrices, even when the interactions between the relevant degrees of freedom  are   not known.
 Sparse non-Hermitian random matrices  generalize  random-matrix ensembles  with independent and identical distributed matrix elements  \cite{ginibre1965statistical, girko1985elliptic, PhysRevLett.60.1895, feinberg1997non, feinberg1997nonx, fyodorov1997almost, janik1997non, akemann2011oxford, bordenave2012around}.      A general theory has been developed  for the spectral density  of  sparse and non-Hermitian random matrices     \cite{Fyodorov, Rogers2,  Metz, Metz2,  Neri2012, saade2014spectral, rouault2015spectrum, amir2016non},  but   other  spectral properties of these ensembles are still poorly understood.

Of particular importance  are  {\it eigenvalue outliers}, which are isolated eigenvalues located outside the continuous (bulk) part of the spectrum (see Fig.~1(a)).  
Eigenvalue outliers of  sparse non-Hermitian random-matrix ensembles, and their associated eigenvectors, are  of key interest for studies on the dynamics of complex systems, and for the evaluation of  ranking and inference algorithms on graphs.     The stationary state of a stochastic process is given by the  left eigenvector associated to an outlier of a Markov matrix, the relaxation time  is given by the corresponding spectral gap  \cite{levin2009markov, monthus}, and the large-deviation function of an observable  is given by an outlier of a modified Markov matrix \cite{donsker1975asymptoticI, donsker1975asymptoticII, donsker1976asymptoticIII, donsker1983asymptoticIV, de2016rare}.   
   Complex dynamical systems, such as, neural networks  \cite{sompolinsky1988chaos, rajan2006eigenvalue, ahmadian2015properties, aljadeff2015transition} or ecosystems \cite{may1972will, allesina2015stability}, are often modelled in terms of differential equations coupled through random matrices.   The eigenvalue with the largest real part, which is often an outlier,  determines
the local stability of these systems \cite{FootnoteF, Fyodorov2016}.   
The PageRank algorithm of Google Search ranks pages of the World Wide Web
with the eigenvector associated to the outlier of a generator matrix
of a stochastic process \cite{langville2011google, ermann2015google}.    Spectral algorithms  detect communities in sparse graphs based on the     eigenvectors of  outliers in the spectrum of the non-backtracking matrix \cite{krzakala2013spectral, saade2014spectral, bordenave2015non}.  If these outliers exist, then it is possible to detect  communities.  Conversely, if these outliers do not exist, then it is impossible for any algorithm to detect  communities.    Quite apart from these applications, the study of outliers of random matrices is also a topic of interest in
 mathematics  \cite{tao2013outliers, bordenave2014outlier}.

In this Letter we present a general theory  for the  outliers of matrices with a {\it local tree} structure.   
We present a set of exact relations for the outliers of sparse  non-Hermitian random matrices, and for the left- and right-eigenvector elements associated to the outlier.    For  oriented random matrices or oriented random graphs, i.e., directed graphs that have no bidirected links, we present explicit expressions  for the eigenvalue outliers, the spectral gap, and the first two moments of the distribution of eigenvector elements  associated to the outlier.  Interestingly, we show that the eigenvalue outliers of oriented random matrices,  and the associated eigenvector moments,  obey universal expressions.

\paragraph{Outliers of non-Hermitian matrices}
We consider an $n\times n$ random matrix $\bA_n$ with probability density $p(\bA_n)$. The matrix $\bA_n$ has $n$ complex-valued eigenvalues $\lambda_1, \cdots, \lambda_n$, and its empirical spectral distribution  is \cite{tao2012topics}:
\begin{eqnarray}
\mu_{\bA_n} =  \frac{1}{n}\sum^n_{j=1}\delta_{\lambda_j}\,,
\end{eqnarray}
with $\delta_{\lambda_j}$ the Dirac measure, i.e.,~$\delta_{\lambda_j}(S)=0$ when $\lambda_j\notin S$ and $\delta_{\lambda_j}(S)=1$ when $\lambda_j\in S$, with $S$ a Lebesgue-measurable subset of $\mathbb{C}$. We assume that the matrix ensembles considered here are self-averaging, i.e., $\mu_{\bA_n}\rightarrow \mu$  for $n\rightarrow \infty$, with $\mu$ a deterministic measure. The Lebesgue decomposition theorem \cite{hewitt2013real} states that $\mu$ consists of an absolute continuous part $\mu_{\rm ac}$, a singular continuous part $\mu_{\rm sing}$, and a pure point part $\mu_{\rm pp}$.
The spectral density function $\rho(\lambda)$, also called the density of states,  is the probability-density function of  $\mu_{\rm ac}$ \cite{spec}. Its support is the subset $\Omega$ of $\mathbb{C}$ for which $\rho(\lambda)>0$, and $\partial \Omega$ is the boundary of $\Omega$. 
The measure $\mu_{\rm pp}$ is discrete, i.e.,  it consists of a  collection of Dirac measures,  $\mu_{\rm pp} = \sum_{\alpha\in \mathcal{L}}a_{\alpha}\delta_{\lambda_{\alpha}}$, where $\mathcal{L}$ defines a countable set
and $a_{\alpha}$ is the weight of the eigenvalue $\lambda_\alpha$.   The outliers of a random matrix  are the values of  $\lambda_{\alpha}$ that lie outside the support of the spectral density $\Omega$ ($\lambda_{\alpha}\notin\Omega$).     In Fig.~1(a) we show for a random matrix the outlier $\lambda_{\rm isol}$ and the boundary $\partial \Omega$ of the support of the spectral density.   

\paragraph{Sparse matrices}
We consider a sparse random and non-Hermitian matrix $\bA_n$.   The matrix elements of $\bA_n$ are 
 $\left[\bA_n\right]_{jk} = C_{jk}J_{jk}$, with $C_{jk}$ the   elements of the adjacency matrix of a random and directed graph \cite{bollobas1998random}, and $J_{jk}$  complex-valued weights that determine  the dynamics of a process on  a graph.  A connectivity element $C_{jk}$ is equal to either $0$ or $1$; if there is a directed link from vertex $j$ to vertex $k$,  then $C_{jk}=1$, whereas if there is no link between the two vertices, then $C_{jk}=0$; we set diagonal elements $C_{ii}$ to one.
We consider graph ensembles  of finite connectivity, in other words, the outdegrees $K^{\rm out}_j = \sum^n_{k=1(k\neq j)}C_{jk}$ and the indegrees $K^{\rm in}_j = \sum^n_{k=1(k\neq j)}C_{kj}$ are finite and independent of $n$.  
Additionally, we consider that the random graph with adjacency matrix $C_{jk}$ is locally tree-like \cite{Bor2010}, which means that a typical neighbourhood of a vertex has  no cycles of degree three or higher \cite{local}.  
Examples of  local tree-like ensembles are  the regular directed graph \cite{Metz, Neri2012}, and the  directed Erd\"os-R\'enyi or Poisson ensemble \cite{Rogers2}.

\paragraph{General theory}
We present  a theory for  the  outliers $\lambda_{{\rm isol}}$ of  locally tree-like random matrices $\bA_n$, and their corresponding left and right eigenvectors, which we denote 
by  $\langle l_{{\rm isol}} \, |$ and $| \, r_{{\rm isol}} \rangle$, respectively.   We first write the right and left eigenvectors of a given outlier $\lambda_{\rm isol}$ in terms of the resolvent  $\bG_n$ of a matrix $\bA_n$.   We define the resolvent $\bG_n(\lambda)$ of the matrix $\bA_n$ as
\begin{eqnarray}
\bG_n(\lambda) \equiv  \left(\bA_n - \lambda \bone_n\right)^{-1}\,, 
\end{eqnarray} 
with $\lambda\in\mathbb{C}$.
The resolvent $\bG_n$  is singular at the eigenvalues $\lambda = \lambda_j$ of $\bA_n$.
Indeed, when we apply the  eigen-decomposition theorem  to $\bG_n$, we find
\begin{eqnarray}
\bG_n =  \frac{| \, r_{\rm isol} \rangle \langle l_{\rm isol} \,|}{\lambda_{\rm isol}-\lambda}  + \sum^n_{j=2} \frac{| \, v^{(r)}_{j} \rangle \langle v^{(l)}_{j} \,| }{\lambda_{j}-\lambda}\,,
\end{eqnarray}
with $| \, v^{(r)}_{j} \rangle$ and  $\langle v^{(l)}_{j} \,|$, respectively, the right and left eigenvectors associated to $\lambda_j$.
If we set $\lambda = \lambda_{\rm isol} - i\eta$, with $\eta$ a small real-valued regularizer, then we have
\begin{eqnarray}
\lim_{\eta\rightarrow 0} i\eta\: \bG_n(\lambda_{\rm isol}-i\eta) =  | \, r_{\rm isol} \rangle \langle l_{\rm isol} \,| + \mathcal{O}\left(\eta\right)\,. \label{eq:res}
\end{eqnarray}
Since $\lambda_{\rm isol}$ is an outlier, the relation (\ref{eq:res})  holds, and is well defined in the infinite-size limit $n\rightarrow \infty$.  

We compute the elements of the resolvent $\bG_n(\lambda-i\eta)$  using the local tree structure  of sparse ensembles in the infinite-size limit. The outcome of our procedure is  a set of recursive equations for the eigenvector elements $r_j =  \langle j | \, r_{\rm isol} \rangle$ and $l_j  =  \langle j | \, l_{\rm isol} \rangle $ (see Supplement \cite{Supp}):
\begin{eqnarray}
  r_j &=&-g_j\sum_{k\in \partial_j } A_{jk}r^{(j)}_k \,, \label{eq:r1} \\ 
l_j^\ast&=& -g_j\sum_{k\in \partial_j}\left(l^{(j)}_k\right)^\ast A_{kj} \label{eq:l1}\,,
\end{eqnarray}
with the "neighbourhood" $\partial_j$ the set of  vertices $k(\neq j)$ for which either $C_{kj}\neq 0$ or $C_{jk}\neq 0$.
The variables $g_j$ are the diagonal elements of the resolvent $\bG_n$, i.e., $g_j = \left[\bG_n(\lambda-i\eta)\right]_{jj}$.  They solve the equations
\begin{eqnarray}
g_j &=&\frac{1}{-\lambda+i\:\eta +A_{jj}- \sum_{k\in\partial_i}A_{jk}\, g^{(j)}_k A_{kj}} \label{eq:res1} \,,\\ 
g^{(\ell)}_j &=&\frac{1}{-\lambda+i\:\eta +A_{jj}- \sum_{k\in\partial_i\setminus \left\{\ell\right\}}A_{jk}\,g^{(j)}_k A_{kj}}\label{eq:res2}\,,
\end{eqnarray} 
for $\lambda\notin \Omega$.    The random variables   $r^{(\ell)}_j$ and $l^{(\ell)}_j$ in Eqs.~(\ref{eq:r1})-(\ref{eq:l1}) solve
\begin{eqnarray} 
  r^{(\ell)}_j &=&-g^{(\ell)}_j\sum_{k\in \partial_j \setminus\left\{\ell\right\}} A_{jk}r^{(j)}_k \,, \label{eq:r1xx} \\ 
\left(l^{(\ell)}_j\right)^\ast&=&-g^{(\ell)}_j\sum_{k\in \partial_j\setminus\left\{\ell\right\}}\left(l^{(j)}_k\right)^\ast A_{kj} \label{eq:l1xx}\,,
\end{eqnarray}
with $\ell\in \partial_j$. 
An outlier value $\lambda_{\rm isol}$ is given by a value $\lambda$ for which the Eqs.~(\ref{eq:r1})-(\ref{eq:l1xx})  admit a non-trivial solution, i.e., a solution for which all eigenvector components $r_j$ and $l_j^\ast$  are neither  zero-valued  nor infinitely large.     The Eqs.~(\ref{eq:r1})-(\ref{eq:l1xx}) apply to non-Hermitian matrices  with a local tree  structure, and extend  studies on the largest eigenvalue of sparse symmetric matrices  \cite{Kab2010, Kab2012, Tak2014, Kaw2015}.

\paragraph{Oriented matrices}
We illustrate our theory on oriented random-matrix ensembles. Oriented matrices  contain only directed links, i.e., $C_{jk}C_{kj} = 0$ for all   $j\neq k$.   For oriented matrices the resolvent Eqs.~(\ref{eq:res1}) and (\ref{eq:res2}) simplify and admit the   solution
\begin{eqnarray}
g_j = g^{(\ell)}_{j} = \left(-\lambda+A_{jj}
\right)^{-1}\quad. \label{eq:gi}
\end{eqnarray}
  The eigenvector components are then given by
\begin{eqnarray}
r_j &=& r^{(\ell)}_{j},\ {\rm for \ all},\  \ell\in \partial^{\rm in}_{j}\, , \label{eqx} \\
l_j &=& l^{(\ell)}_{j},\ {\rm for \ all},\  \ell\in \partial^{\rm out}_{j}\, ,  \label{eqxx}
\end{eqnarray}
where the random variables $r^{(\ell)}_{j}$ and  $l^{(\ell)}_{j}$ represent a non-trivial solution to the Eqs.~(\ref{eq:r1xx})-(\ref{eq:l1xx}).   The "in-neighbourhood" $\partial^{\rm in}_{j}$ is the set of vertices $k$($\neq j$) with $C_{kj}\neq 0$, and  the  "out-neighbourhood" $\partial^{\rm out}_{j}$ is the set of vertices $k$($\neq j$) w§ith $C_{jk}\neq 0$.

\begin{figure}[h]
 \begin{center}
 \includegraphics[width=0.5
\textwidth]{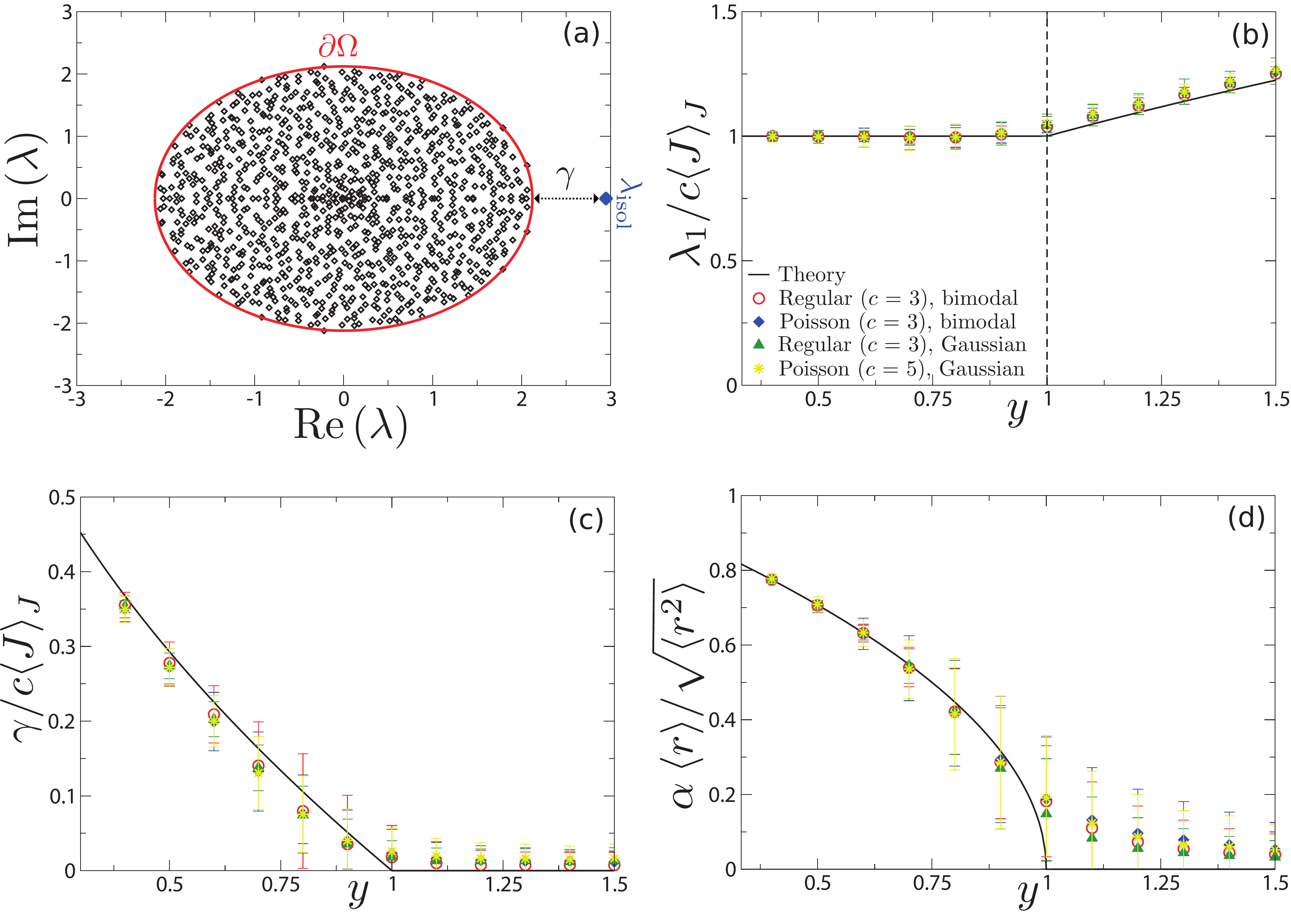}
 \caption{The outlier $\lambda_{\rm isol}$, spectral gap $\gamma$ and first moment $\langle r \rangle$ of the eigenvector, associated to $\lambda_{\rm isol}$, of  oriented adjacency matrices.   Direct-diagonalization results of  matrices of finite size $n=1000$ (markers) are compared with our theory for infinite-sized matrices,  given by Eqs.~(\ref{eq:outlier})-(\ref{jsos})  (solid lines). 
{\it Subfigure (a):} eigenvalues of one  $c$-regular matrix with Gaussian distributed off-diagonal elements, mean degree $c=3$ and $y = 0.5$, with $y = \langle J^2\rangle_J/(c\langle J \rangle_J^2)$ 
 the disorder parameter.  The boundary  $\partial \Omega$ of the support of the spectral density, the spectral gap $\gamma$ and the outlier $\lambda_{\rm isol}$ are indicated.   
{\it Subfigures (b)-(d):} the eigenvalue $\lambda_{\rm 1}$ with the largest real part, the spectral gap $\gamma$, and  the first moment $\langle r \rangle$  of the right eigenvector associated to $\lambda_1$, all plotted  as a function of $y$.  The eigenvalue $\lambda_1$ is an outlier for  $y<1$, i.e., $\lambda_1=\lambda_{\rm isol}$, and $\lambda_1\in \partial \Omega$ for $y>1$.   Results shown are  for  four different  ensembles of oriented matrices.   The ensembles are either $c$-regular or  Poissonian with mean connectivity $c$; nonzero off-diagonal elements are i.i.d.~with either a bimodal distribution $p_{\rm J}(J) =(1-\Delta)\delta(J+1) + \Delta \delta(J-1)$, or  a Gaussian
distribution with mean $\langle J \rangle_J = 1$; diagonal matrix elements are set to zero.   Direct diagonalization results in subfigures (b)-(d) are from $1000$ samples.   Error bars represent the  standard deviation of the sampled population and $\alpha = \sqrt{\langle (K^{\rm out})^2\rangle_{K^{\rm out}}-c }/c$.
}\label{fig:1}
 \end{center}
\end{figure}

\begin{figure}[h]
 \begin{center}
 \includegraphics[width=0.5
\textwidth]{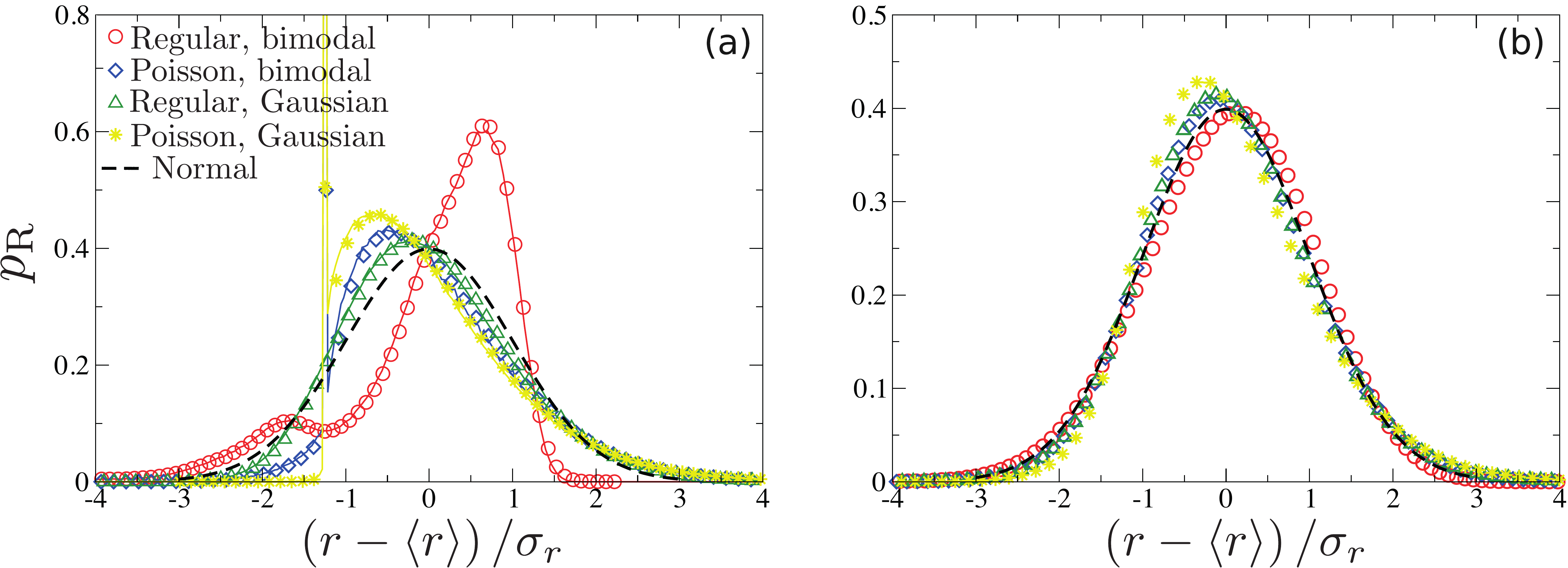}
 \caption{Probability distribution $p_{\rm R}$ of the right  eigenvector elements associated to the outlier of oriented adjacency matrices.   The ensembles are the same in Fig.1 with disorder parameter $y=0.4$, and   mean connectivity  (a) $c=3$,  or (b) $c=10$  .    We compare  direct-diagonalization results (markers) with population-dynamics results (solid lines in  (a)) and with the normal distribution (dashed line).    Direct-diagonalization results are for  $2e+4$ matrix samples of size $n=1000$.   In order to show  universality of the distribution at high connectivities, we have rescaled the distributions with their mean $\langle r \rangle$ and  standard deviation $\sigma_r$. 
}\label{fig:2}
 \end{center}
\end{figure}

We derive explicit analytical and numerical results by   ensemble averaging the Eqs.~(\ref{eqx})-(\ref{eqxx}).  An outlier value $\lambda_{\rm isol}$, and its associated eigenvector moments $\langle r^m\rangle =  N^{-1}\langle \sum^n_{j=1} r^m_j\rangle $ and 
$\langle l^m\rangle =  n^{-1}\langle \sum^n_{j=1} l^m_j\rangle $, with $m=1,2$, are given by a non-trivial solution to  these ensemble-averaged  equations;   the symbol $\langle \dots \rangle$ denotes here the ensemble average with respect to the distribution $p(\bA_n)$.  
Additionally, we can compute the associated ensemble-averaged distribution of eigenvector elements  using the population dynamics algorithm \cite{abou1973selfconsistent, cizeau1994theory, mezard2001bethe, metz2010localization, Supp}.  We illustrate  this ensemble-averaging procedure  on two paradigmatic examples of sparse matrix ensembles:  adjacency matrices and  Laplacian matrices of oriented random graphs.

\paragraph{Adjacency matrices}

We consider random adjacency  matrices 
$\bA_n$, which represent an oriented random graph with a given joint  distribution $p_{K^{\rm in},K^{\rm out}}$  of  in- and outdegrees  \cite{bollobas1998random, molloy1995critical, molloy1998size}.   The  off-diagonal weights $J_{kj}$, with $k\neq j$,   are independent and identically   distributed (i.i.d.) with  distribution $p_{\rm J}$, and  the diagonal weights $J_{jj}$  are
i.i.d.~with distribution $p_{\rm D}$.   

The oriented adjacency matrices  we consider here have either exactly one outlier (see Fig.~\ref{fig:1}(a)), or do not have any outlier.    
If the outlier exists,  we call the random-matrix ensemble  gapped.  Conversely,   if  the outlier does not exist,  we call the ensemble gapless.  If the outlier exists, its 
value $\lambda_{\rm isol}$  solves  \cite{Supp}
\begin{eqnarray}
\Big\langle \left(\lambda_{\rm isol}-D\right)^{-1}\Big\rangle_{\rm D} = \frac{1}{c\: \langle J \rangle_{\rm J}} \,  ,  \label{eq:outlier}
\end{eqnarray}
with  $\langle \cdot \rangle_D$ and $\langle \cdot \rangle_J$ denoting, respectively, the 
average with respect to the distributions $p_D$ and $p_J$.    The quantity $c= \langle K^{\rm in}\rangle_{K^{\rm in}} = \langle K^{\rm out} \rangle_{K^{\rm out}}$ 
is the mean degree of the graph, where $\langle \cdot \rangle_{K^{\rm in}}$ and $\langle \cdot \rangle_{K^{\rm out}}$ denote averages with respect to the indegree and outdegree distribution, respectively.  Equation (\ref{eq:outlier}) follows from solving the ensemble averaged version of the  Eqs.~(\ref{eq:r1}) and (\ref{eq:l1}) for the eigenvector moments.    The first two moments of the distribution of right- and left-eigenvector elements  read \cite{Supp}
\begin{eqnarray}
\langle r\rangle^2/\langle r^2\rangle  &=& \frac{\mathcal{Q}}{  \left\langle (K^{\rm out})^{2}   \right\rangle_{K_{\rm out}}  - c } \, , \\
\langle l\rangle^{2}/\langle l^2\rangle    &=& \frac{\mathcal{Q}}{  \left\langle (K^{\rm in})^{2}   \right\rangle_{K_{\rm in}}   - c   } \, ,  \label{Eq:l}
\end{eqnarray} 
with  $\mathcal{Q} =   \Big\langle \langle J  \rangle_{\rm J} ^2/|\lambda_{\rm isol}- D|^2 \Big\rangle^{-1}_{\rm D} -  c \langle J^2 \rangle_{\rm J}/\langle J  \rangle^2_{\rm J} $.     Additionally, we find  
the support $\Omega$ of $\rho(\lambda)$ from a stability analysis around the solution (\ref{eq:gi}) to the resolvent Eqs.~(\ref{eq:res1}) and (\ref{eq:res2});  the set   $\Omega$ contains the values $\lambda\in\mathbb{C}$ with
\begin{equation}
 \left\langle  \frac{1 }{|\lambda - D|^2}   \right\rangle^{-1}_D <  c \left\langle J^2 \right\rangle_J\,.
\label{jsos}
\end{equation}   
In Fig.~\ref{fig:1}  we compare the analytical expressions, given by Eqs.~(\ref{eq:outlier})-(\ref{jsos}), with direct-diagonalization results of matrices  of  finite size.  Results are in good correspondence and 
converge to the theoretical expressions for large matrix sizes $n\gg 1$ (for which the ensembles become local tree like).

Equations (\ref{eq:outlier})-(\ref{jsos})  imply that the outlier of oriented adjacency matrices, and the  first moments of its associated eigenvector distribution, are universal.     In order to illustrate this universality, we plot in Fig.~\ref{fig:1}(b)-(d), for different matrix ensembles,  the eigenvalue outlier, the spectral gap and the first two moments of the eigenvector distribution, as a function of the disorder parameter $y = \langle J^2\rangle_J/\left( c\langle J \rangle^2_J \right)$.  The curves for the different ensembles collapse on the  universal curve  given by our analytical expressions Eqs.~(\ref{eq:outlier}-\ref{jsos}). 

 A characteristic feature of Fig.~\ref{fig:1} is the phase transition from a  gapless phase at high disorder, $y>1$,  to a gapped phase at low disorder, $y<1$. Note that this phase transition is generic and it also appears in symmetric random matrix ensembles \cite{Edwards, Bassler,Kab2010, Kab2012, Tak2014, Kaw2015}. 
 
For large mean connectivities, $c \gg 1$, the distributions of  right- and left-eigenvector elements become universal, and from Eqs.~(\ref{eq:r1}) and (\ref{eq:l1}), it follows that they are Gaussian.  In Fig.~\ref{fig:2}(b) we illustrate the universality of the eigenvector distributions at high connectivities $c$.   At low connectivities $c$, the distributions are not universal, but direct-diagonalization results are in good correspondence with numerical solutions of Eqs.~(\ref{eq:r1}) and (\ref{eq:l1}) using the population dynamics algorithm  (see Fig.~\ref{fig:2}(a)).

\paragraph{Laplacian matrices}
Laplacian matrices are the generator matrices of the dynamics of a random walk on a graph.     
The defining feature of a Laplacian matrix is the constraint $J_{jj} = -\sum^n_{k=1, (k\neq j)} J_{jk}$ on its diagonal elements.    Symmetric Laplacian matrices have been studied in~\cite{Fyodorov2003, Kuhn2015}.
Here we study the spectra of unnormalized Laplacian matrices of oriented graphs with off-diagonal matrix elements  $J_{jk} = 1$ and with a  given joint degree  distribution $p_{K^{\rm in},K^{\rm out}}$   \cite{bollobas1998random, molloy1995critical, molloy1998size}.

\begin{figure}[h]
 \begin{center}
 \includegraphics[width=0.5 \textwidth]{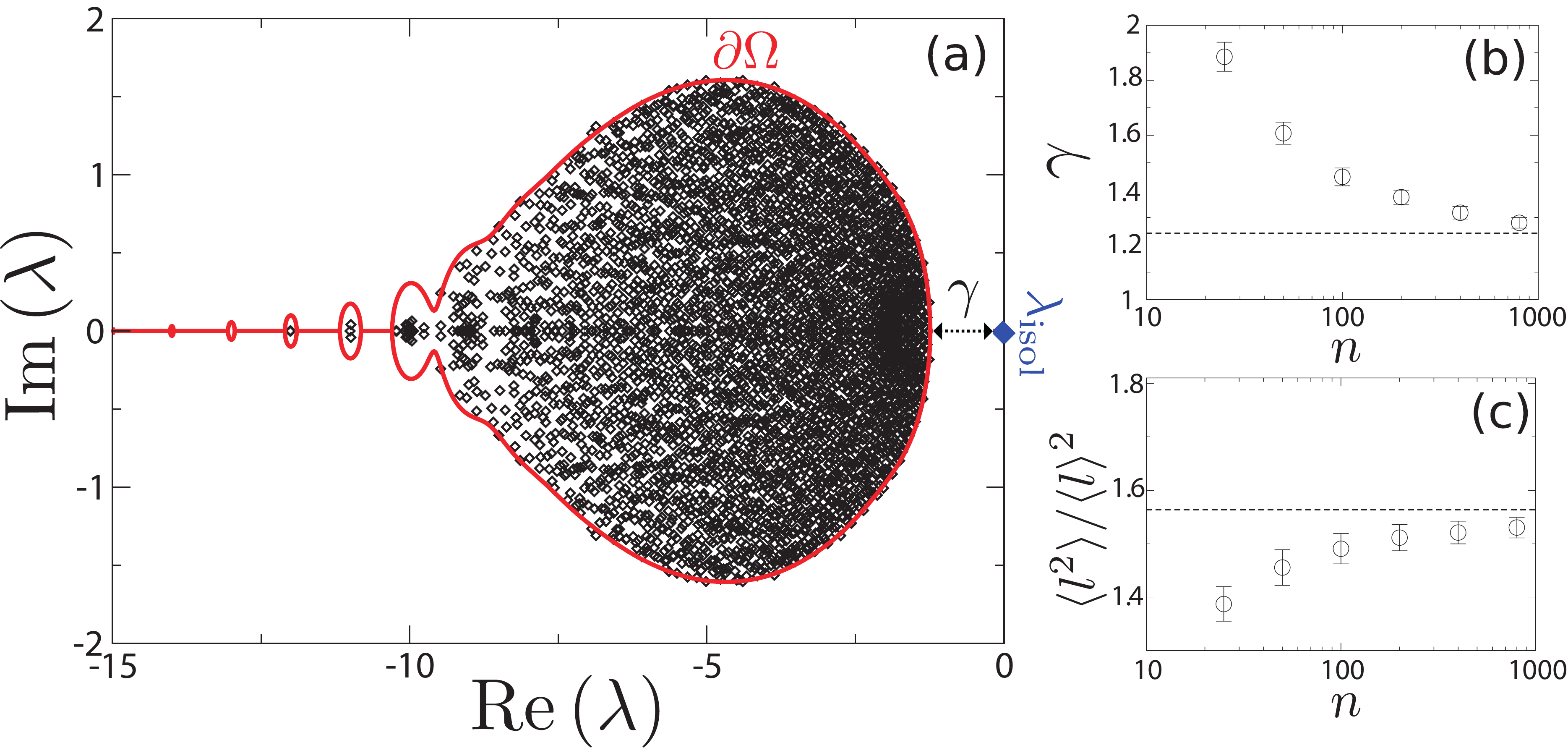}
 \caption{Results for  Laplacian matrices defined on an oriented Erd\"os-R\'enyi random graph with off-diagonal matrix elements 
 $J_{kj}=1$ for $k\neq j$.  We consider here an ensemble with correlated in- and outdegrees: 
 $p_{K^{\rm in}, K^{\rm out}}(k^{\rm in}, k^{\rm out}) = \delta(k^{\rm in};k)\delta(k^{\rm out};k)p_{\rm deg}(k)$.    The degree distribution $p_{\rm deg}(k)$ is Poissonian, i.e.,
 $p_{\rm deg}(k) = \mathcal{N}e^{-\tilde{c}}\tilde{c}^k/k!$, if $k\geq k_0$, and $p_{\rm deg}(k) =0$  if $k<k_0$.  
 Direct diagonalization results (markers) are compared with analytical results (solid lines) for $k_0 = 2$ and   $\tilde{c}=4$.
{\it Subfigure}~(a): spectrum of a single matrix with $n=4000$.  The red line shows the  boundary of the support of the spectral density $\partial \Omega$, given by Eqs.~(\ref{eq:supp1}). {\it Subfigures} ~(b)-(c): spectral gap $\gamma$ and 
 moments $\langle l^{2} \rangle/\langle l \rangle^2$ are shown to converge to the theoretical values for $n\rightarrow \infty$.     Direct diagonalization results are averages over $1e+3$ matrices (markers) and theoretical expressions follow from Eqs.~(\ref{eq:second})-(\ref{eq:supp1}) (dashed lines). 
}\label{fig:3}
 \end{center}
\end{figure}

The outlier of  Laplacian matrices is given by  $\lambda_{\rm isol} = 0$, and the distribution of right-eigenvector elements reads $p_{\rm R}(r) = \delta(r-1)$.  The distribution of left-eigenvector elements $p_{\rm L}(l)$ encodes the statistics of the  steady-state probability distribution of a random walk on the associated graph. 
We take the average of Eqs.~(\ref{eq:l1}) and find for the moments of the distribution of left-eigenvector elements (see Supplement \cite{Supp}):
 \begin{equation}
  \frac{\langle l^2\rangle}{\langle l\rangle^2} = 
\frac{\left\langle  \frac{(K^{\rm in})^2-c}{(K^{\rm out})^2}   \right\rangle_{K^{\rm in}, K^{\rm out}} }
{\left\langle  \frac{c}{  K^{\rm out}    }  \right\rangle^{2}_{K^{\rm out}}- \left\langle  \frac{c }{ (K^{\rm out})^2}   \right\rangle_{K^{\rm out}}  } \,.
 \label{eq:second}
\end{equation} 
   We also derive an expression for the  support $\Omega$ of the spectral-density function.   We find that  $\Omega$ consists of values $\lambda\in\mathbb{C}$ for which  either 
\begin{eqnarray}
 \Bigg\langle\frac{K^{\rm out}}{\left|\lambda+K^{\rm out}\right|^2}  \Bigg\rangle_{K^{\rm out}}> 1\,, {\rm or} \ \Bigg\langle\frac{K^{\rm in}}{\left|\lambda+K^{\rm out}\right|^2}  \Bigg\rangle_{K^{\rm in}, K^{\rm out}}> 1\,. \nonumber\\ \label{eq:supp1}
\end{eqnarray} 
In Fig.~\ref{fig:3}(a) we compare the Eqs.~(\ref{eq:supp1})  for $\Omega$ with direct-diagonalization results of Laplacian matrices of finite size.  
We also compare direct-diagonalization results for the spectral gap $\gamma$ and the ratio of the moments  $\langle l^2\rangle/\langle l\rangle^{2}$ with the exact expressions given by Eqs.~(\ref{eq:second})-(\ref{eq:supp1}).      Numerical results converge to the analytical expressions for large  matrix sizes~$n$.

\paragraph{Discussion}  
We have presented an exact  theory for the outliers of random matrices with a local tree structure.  Remarkably, for oriented matrices we find general analytical expressions for the outliers, the associated statistics of  eigenvectors, and the support of the spectral density.  These results show that the spectral properties of outliers of oriented matrices are universal.   It will be interesting to explore the implications of these results for the dynamics of complex systems with unidirectional interactions, which often appear in biological systems that operate far from thermal equilibrium, for example,  neural networks \cite{amit1997dynamics, brunel2000dynamics} or  networks of biochemical reactions \cite{edwards2000escherichia}.   Our theory, based on the Eqs.~(\ref{eq:r1})-(\ref{eq:l1xx}), applies also to non-oriented matrices, and we illustrate this on the elliptic regular ensemble in the supplement \cite{Supp}.     Following Refs.~\cite{Metz, Metz2} it is  possible to extend our approach   
to random matrices with many cycles.   We expect  that studies  along these lines will  lead to  a general theory for  the outliers of sparse random matrices.    

I.N. thanks Jos\'{e} Negrete Jr. for a stimulating discussion.

\clearpage
\setboolean{@twoside}{false}


\section*{Supplementary material}

# Supplemental Material: Eigenvalue Outliers of non-Hermitian Random Matrices with a Local Tree Structure

Izaak Neri[1,2] and Fernando Lucas Metz[3]

[1] Max Planck Institute for the Physics of Complex Systems, Nöthnitzerstraße 38, 01187 Dresden, Germany.
[2] Max Planck Institute of Molecular Cell Biology and Genetics,
Pfotenhauerstraße 108, 01307 Dresden, Germany.
[3] Departamento de Física, Universidade Federal de Santa Maria, 97105-900 Santa Maria, Brazil.

## S1. INTRODUCTION

In this Supplement we derive the Eqs. (5)-(10) yielding the isolated eigenvalue, and its associated left and right eigenvectors, of sparse non-Hermitian random matrices. Additionally, we derive the analytical expressions, given by Eqs. (15)-(20), for the outlier, the moments of the associated left and right eigenvectors, and the spectral gap of adjacency and Laplacian matrices of oriented graphs. We end the supplement with an illustration of our theory on a non-oriented sparse random matrix ensemble.

The Supplement is structured as follow. In section S2, we discuss some definitions on eigenvalues of random matrices. In section S3, we introduce the Hermitization method, which is a regularization method for non-Hermitian random matrices. In section S4, we first formulate the eigenvalue-outlier problem as a matrix inversion problem, we then show how this problem can be solved using the local tree structure of sparse matrices, and we finally derive the main Eqs. (5)-(10). In section S5, we illustrate our theory on oriented matrices with a local tree structure. In the following two sections S6 and S7, we apply our theory on oriented adjacency and Laplacian matrices, respectively, and derive the Eqs. (15)-(20). In the last section S8, we illustrate our theory on a non-oriented random matrix ensemble.

## S2. PRELIMINARIES

We consider a square non-Hermitian random matrix $\boldsymbol{A}_n$ of size $n$ with matrix elements $A_{k\ell} = (\boldsymbol{A}_n)_{k\ell} = J_{k\ell} C_{k\ell}$. The random matrix $\boldsymbol{A}_n$ is sparse: the variables $C_{k\ell}$ are the matrix elements of the adjacency matrix of a sparse directed graph, and the complex-valued variables $J_{k\ell}$ determine the interactions between the degrees of freedom of a process on this graph (see main text).

The matrix $\boldsymbol{A}_n$ has $n$ complex-valued eigenvalues $\lambda_k(\boldsymbol{A}_n)$, which are the $n$ roots of the characteristic polynomial $\det(\boldsymbol{A}_n - \lambda \boldsymbol{1}_n) = 0$. We label the eigenvalues $\lambda_1(\boldsymbol{A}_n), \lambda_2(\boldsymbol{A}_n), \ldots, \lambda_n(\boldsymbol{A}_n)$ with indices such that $|\lambda_1(\boldsymbol{A}_n)| \geq |\lambda_2(\boldsymbol{A}_n)| \geq \ldots |\lambda_n(\boldsymbol{A}_n)|$; if $|\lambda_j| = |\lambda_{j+1}|$ we label the eigenvalues such that the real part of $\lambda_j$ is larger than the real part of $\lambda_{j+1}$. We write the right and left eigenvectors associated to the eigenvalue $\lambda_j$ as $|v_j^{(r)}\rangle$ and $\langle v_j^{(l)}|$, respectively.

We call $\lambda_j$ an *outlier* if, with probability one, $\lambda_j \notin \Omega$ for $n \rightarrow \infty$. Here, the set $\Omega$ is the support of the spectral density, i.e., the set of all values $\lambda \in \mathbb{C}$ for which the spectral density function $\rho(\lambda) > 0$. The spectral density $\rho(\lambda)$ is the absolute continuous part of the empirical spectral distribution in the limit $n \rightarrow \infty$ (see main text). In the examples we consider, the outlier is always given by $\lambda_1$ (if it exists), and we write $\lambda_{\rm isol} = \lambda_1$, $|r_{\rm isol}\rangle = |v_1^{(r)}\rangle$ and $\langle l_{\rm isol}| = \langle v_1^{(l)}|$.

We also define the singular values $\{s_j\}_{j=1..n}$ of $\boldsymbol{A}_n$, which are the eigenvalues of the matrix $\sqrt{\boldsymbol{A}_n \boldsymbol{A}_n^\dagger}$, i.e., $s_j(\boldsymbol{A}_n) = \lambda_j\left(\sqrt{\boldsymbol{A}_n \boldsymbol{A}_n^\dagger}\right) > 0$. The singular values are also the eigenvalues of the Hermitized matrix

$$\boldsymbol{H}_{2n} = \begin{pmatrix} 0_n & \boldsymbol{A}_n \\ \boldsymbol{A}_n^\dagger & 0_n \end{pmatrix}, \tag{S1}$$

which has the eigenvalues $\pm s_j(\boldsymbol{A}_n)$. We denote conjugate-transposition by the symbol "$\dagger$", and we denote complex conjugation by the symbol "$*$".

## S3. HERMITIZATION METHOD FOR NON-HERMITIAN MATRIX ENSEMBLES

We revisit the Hermitization method [1, 2]. The Hermitization method regularizes the resolvent $\boldsymbol{G}_n = (\boldsymbol{A}_n - \lambda \boldsymbol{1}_n)^{-1}$ of a matrix $\boldsymbol{A}_n$, in other words, it removes its singular points from its domain. Through Hermitization we can work

directly in the asymptotic limit $n \to \infty$, albeit using a regularized ensemble. Working in the asymptotic limit $n \to \infty$ is of advantage, since sparse non-Hermitian ensembles are local tree like in this limit, and therefore become tractable with recursive methods [3, 4].

We regularize the ensemble $\boldsymbol{A}_n$ by defining the normal matrix $\boldsymbol{M}_{2n}(\lambda, \eta)$ [2–4]:

$$\boldsymbol{M}_{2n}(\lambda, \eta) = \begin{pmatrix} \eta \, \boldsymbol{1}_n & -i \left( \boldsymbol{A}_n - \lambda \, \boldsymbol{1}_n \right) \\ -i \left( \boldsymbol{A}_n^\dagger - \lambda^* \, \boldsymbol{1}_n \right) & \eta \, \boldsymbol{1}_n \end{pmatrix}. \tag{S2}$$

Since $\boldsymbol{M}_{2n}^\dagger \boldsymbol{M}_{2n} = \boldsymbol{M}_{2n} \boldsymbol{M}_{2n}^\dagger$, the matrix $\boldsymbol{M}_{2n}$ is unitarily similar to a diagonal matrix. The matrix $\boldsymbol{M}_{2n}$ can also be written as

$$\boldsymbol{M}_{2n}(\lambda, \eta) = \eta \, \boldsymbol{1}_{2n} - i \, \boldsymbol{B}_{2n}(\lambda), \tag{S3}$$

with $\boldsymbol{B}_{2n}$ a Hermitian matrix. The Hermitian matrix $\boldsymbol{B}_{2n}(\lambda, \eta)$ is the same as in Eq.(2) of Ref. [4], and appears also in the work [3]. In these references the spectral density function $\rho(\lambda) = -\lim_{n \to \infty} (n \, \pi)^{-1} \partial^* \text{Tr} \, \boldsymbol{G}_n$ of sparse non-Hermitian ensembles is computed using the regularized random matrix $\boldsymbol{B}_{2n}(\lambda, \eta)$ ("Tr" is the trace operator and $\partial^* = 1/2(\partial_x + i\partial_y)$).

The inverse of the matrix $\boldsymbol{M}_{2n}(\lambda, \eta)$ is:

$$-i \boldsymbol{M}_{2n}^{-1}(\lambda, \eta) = \begin{pmatrix} \boldsymbol{C}_n(\lambda, \eta) & \tilde{\boldsymbol{G}}_n^\dagger(\lambda, \eta) \\ \tilde{\boldsymbol{G}}_n(\lambda, \eta) & \boldsymbol{D}_n(\lambda, \eta) \end{pmatrix}, \tag{S4}$$

with

$$\boldsymbol{C}_n = \frac{-i\eta}{\eta^2 \boldsymbol{1}_n + (\boldsymbol{A}_n - \lambda \, \boldsymbol{1}_n)\left(\boldsymbol{A}_n^\dagger - \lambda^* \, \boldsymbol{1}_n\right)}, \tag{S5}$$

$$\begin{aligned}\tilde{\boldsymbol{G}}_n(\lambda, \eta) &= \left(\boldsymbol{A}_n^\dagger - \lambda^* \, \boldsymbol{1}_n\right)\left(-\eta^2 \boldsymbol{1}_n + (\boldsymbol{A}_n - \lambda \, \boldsymbol{1}_n)\left(\boldsymbol{A}_n^\dagger - \lambda^* \, \boldsymbol{1}_n\right)\right)^{-1} \\ &= \boldsymbol{G}_n + \mathcal{O}\left(\eta^{-2}\right),\end{aligned} \tag{S6}$$

$$\begin{aligned}\boldsymbol{D}_n &= -i\eta^{-1}\left[\boldsymbol{1}_n - \left(\boldsymbol{A}_n^\dagger - \lambda^* \, \boldsymbol{1}_n\right)\left(\eta^2 \boldsymbol{1}_n + (\boldsymbol{A}_n - \lambda \, \boldsymbol{1}_n)\left(\boldsymbol{A}_n^\dagger - \lambda^* \, \boldsymbol{1}_n\right)\right)^{-1}(\boldsymbol{A}_n - \lambda \, \boldsymbol{1}_n)\right] \\ &= \boldsymbol{C}_n + \mathcal{O}\left(\eta^{-2}\right).\end{aligned} \tag{S7}$$

The matrix $\tilde{\boldsymbol{G}}_n(\lambda, \eta)$ has no singularities for values $\eta > 0$, and the procedure thus regularizes $\boldsymbol{G}_n(\lambda)$ when $\eta > 0$. This approach is not rigorous, since it is not known whether the two limits $\eta \to 0$ and $n \to \infty$ commute. Nevertheless, comparisons with direct diagonalization results for the spectral density of sparse matrices provide evidence that the method is exact [3–6], which indicates that the two limits commute.

In the following we apply the Hermitization method to the outlier problem of sparse non-Hermitian matrices.

### S4. DERIVATION OF THE EQUATIONS FOR THE OUTLIERS OF LOCAL TREE-LIKE MATRICES

We derive the recursive Eqs. (5)-(10) using the Hermitization method and the local tree-like property of sparse ensembles in the limit $n \to \infty$.

#### A. Outliers and the Hermitization method

We first show how the outliers of sparse non-Hermitian matrix ensembles can be calculated within the framework of the Hermitization method. Indeed, we show that the eigenvectors of an outlier $\lambda_{\text{isol}}$ can be written in terms of the inverse of the normal matrix $\boldsymbol{M}_{2n}(\lambda, \eta)$.

The matrix $\boldsymbol{M}_{2n}(\lambda, \eta)$ has the eigenvalues $\{\eta + i \, s_j(\lambda \boldsymbol{1}_n - \boldsymbol{A}_n)\}_{j=1..n}$ and $\{\eta - i \, s_j(\lambda \boldsymbol{1}_n - \boldsymbol{A}_n)\}_{j=1..n}$. Since $\boldsymbol{M}_{2n}(\lambda, \eta)$ is normal, the right and left eigenvectors associated to these eigenvalues are the same. If we set $\lambda$ equal to an eigenvalue of the matrix $\boldsymbol{A}_n$, for instance $\lambda = \lambda_k(\boldsymbol{A}_n)$, then $\boldsymbol{M}_{2n}(\lambda_k, \eta)$ has the eigenvalue $\eta$ with algebraic multiplicity two.

Let us for now consider that $\lambda_1 = \lambda_{\text{isol}}$, an isolated outlier of the random matrix ensemble. The eigenspace associated to $\eta$ is spanned by the two orthogonal vectors $|v_{\text{isol}}^+\rangle = (|l_{\text{isol}}\rangle, |r_{\text{isol}}\rangle)^T$, and $|v_{\text{isol}}^-\rangle = (-|l_{\text{isol}}\rangle, |r_{\text{isol}}\rangle)^T$. We write $\boldsymbol{M}_{2n}(\lambda_{\text{isol}}, \eta)$ in terms of its eigenvalues and eigenvectors

$$\boldsymbol{M}_{2n}(\lambda_{\text{isol}}, \eta) = \eta \sum_{\sigma=\pm} |v_{\text{isol}}^\sigma\rangle\langle v_{\text{isol}}^\sigma| + \sum_{j=3}^{2n} m_j |v_j\rangle\langle v_j|, \tag{S8}$$

which is possible since $\boldsymbol{M}_{2n}$ is normal. Here, $\{m_j\}_{j=3,\ldots,2n}$ are the $2(n-1)$ remaining eigenvalues of $\boldsymbol{M}_{2n}$, given by $\eta + i\, s_j(\lambda_{\text{isol}} \mathbf{1}_n - \boldsymbol{A}_n)$ and $\eta - i\, s_j(\lambda_{\text{isol}} \mathbf{1}_n - \boldsymbol{A}_n)$, for $j \neq 1$; $|v_j\rangle$ are the corresponding eigenvectors. The inverse of $\boldsymbol{M}_{2n}$ is thus:

$$\boldsymbol{M}_{2n}^{-1} = \eta^{-1} \sum_{\sigma=\pm} |v_{\text{isol}}^\sigma\rangle\langle v_{\text{isol}}^\sigma| + \sum_{j=3}^{2n} m_j^{-1} |v_j\rangle\langle v_j|. \tag{S9}$$

Since $\lambda_{\text{isol}}$ is an outlier, it is isolated from all the other eigenvalues $\lambda_j$ by a gap $\lambda_{\text{isol}} - \lambda_j \sim \mathcal{O}(1)$. Therefore, we have

$$\lim_{\eta \to 0^+} \boldsymbol{M}_{2n}^{-1} = \eta^{-1} \sum_{\sigma=\pm} |v_{\text{isol}}^\sigma\rangle\langle v_{\text{isol}}^\sigma| + \mathcal{O}(1). \tag{S10}$$

We find thus the following expressions for the eigenvectors associated to an outlier $\lambda_{\text{isol}}$

$$\begin{pmatrix} \alpha |l_{\text{isol}}\rangle \\ \beta |r_{\text{isol}}\rangle \end{pmatrix} = \lim_{\eta \to 0^+} \eta\, \boldsymbol{M}_{2n}^{-1}(\lambda_{\text{isol}}, \eta) |1\rangle, \tag{S11}$$

with $\alpha$ and $\beta$ constants, and with $|1\rangle = (1\ 1\ \ldots\ 1)^T$. Without loss of generality we can set the constants $\alpha$ and $\beta$ to one, i.e., $\alpha = 1$ and $\beta = 1$. These constants merely reflect the fact that any scalar multiple of an eigenvector is still an eigenvector.

We have thus shown that the left and right eigenvectors associated to an outlier $\lambda_{\text{isol}}$ can be expressed as the inverse of a matrix, $\lim_{\eta \to 0^+} \eta \boldsymbol{M}_{2n}^{-1}(\lambda_{\text{isol}}, \eta)|1\rangle$. In the next subsection we show how to compute the inverse of $\boldsymbol{M}_{2n}^{-1}$.

### B. Matrix inversion with methods from sparse-random-matrix theory

We show how to compute the vector $\eta \boldsymbol{M}_{2n}^{-1}(\lambda, \eta)|1\rangle$ with methods from sparse-random-matrix theory. We use Gaussian belief propagation to compute the inverse of the normal matrix $\boldsymbol{M}_{2n}^{-1}$ [3, 7–9]; in physics Gaussian belief propagation is often called the cavity method [3, 10]. In the first paragraph, we formulate the inversion of a normal matrix as an inference problem of marginals of a Gaussian distribution. In the second paragraph, we solve this inference problem with Gaussian belief propagation.

#### 1. Matrix inversion and inference in Gaussian distributions

We formulate the matrix-inversion problem $\eta \boldsymbol{M}_{2n}^{-1}(\lambda, \eta)|1\rangle$ as an inference problem on a Gaussian distribution.
Let us therefore introduce the multivariate Gaussian distribution [11]:

$$p_{\hat{\boldsymbol{\phi}}}(\boldsymbol{x}, \boldsymbol{x}^\dagger; \boldsymbol{M}_{2n}) = \frac{1}{\mathcal{Z}_{\hat{\boldsymbol{\phi}}}} \exp\left(-\boldsymbol{x}^\dagger \boldsymbol{M}_{2n} \boldsymbol{x} + i\, \hat{\boldsymbol{\phi}}^\dagger \boldsymbol{x} + i\, \boldsymbol{x}^\dagger \hat{\boldsymbol{\phi}}\right), \tag{S12}$$

$$\mathcal{Z}_{\hat{\boldsymbol{\phi}}} = \frac{(2\pi i)^{2n}}{\det \boldsymbol{M}_{2n}} \exp\left(-\hat{\boldsymbol{\phi}}^\dagger \boldsymbol{M}_{2n}^{-1} \hat{\boldsymbol{\phi}}\right), \tag{S13}$$

where $\boldsymbol{x}$ and $\hat{\boldsymbol{\phi}}$ are $2n$-dimensional vector with complex components, and $p_{\hat{\boldsymbol{\phi}}}$ is a distribution defined on the domain $\boldsymbol{x} \in \mathbb{C}^{2n}$. For simplicity, we write $p_{\hat{\boldsymbol{\phi}}}(\boldsymbol{x}, \boldsymbol{x}^\dagger; \boldsymbol{M}_{2n}) = p_{\hat{\boldsymbol{\phi}}}(\boldsymbol{x}, \boldsymbol{x}^\dagger)$. For $\eta > 0$, the Hermitian part of $\boldsymbol{M}_{2n}$ is positive-definite, which ensures the convergence of integrals involving the distribution of Eq. (S12). Note that this corresponds with the fact that the regularized resolvent $\boldsymbol{G}_n(\lambda, \eta)$ has no singularities for $\eta > 0$, as discussed in Section S3. Moreover,

since $\boldsymbol{M}_{2n}$ is a normal matrix it can be diagonalized by a unitary transformation. Using these two properties, and the formal marginalization of a Gaussian distribution, i.e.,

$$p_{\hat{\boldsymbol{\phi}}}(x_j, x_j^*) = \int \left( \prod_{k=1(k\neq j)}^{2n} dx_k dx_k^* \right) p_{\hat{\boldsymbol{\phi}}}(\boldsymbol{x}, \boldsymbol{x}^\dagger), \tag{S14}$$

we find the identity

$$i \left( \boldsymbol{M}_{2n}^{-1} \hat{\boldsymbol{\phi}} \right)_j = \int dx_j dx_j^* \, x_j \, p_{\hat{\boldsymbol{\phi}}}(x_j, x_j^*). \tag{S15}$$

We apply Eq. (S15) to the vector $\hat{\boldsymbol{\phi}} = \eta |1\rangle$, from which we find an expression for the elements of the vector $\eta \boldsymbol{M}_{2n}^{-1}(\lambda, \eta)|1\rangle$ in terms of the distribution $p_{|1\rangle}(x_j, x_j^*)$, viz.,

$$\eta \langle j | \boldsymbol{M}_{2n}^{-1} | 1 \rangle = -i \int dx_j dx_j^* \, x_j \, p_{|1\rangle}(x_j, x_j^*). \tag{S16}$$

The function $p_{|1\rangle}(x_j, x_j^*)$ is the marginal distribution of the $j$-th vertex of the Gaussian distribution

$$p_{|1\rangle}(\boldsymbol{x}, \boldsymbol{x}^\dagger) = \frac{1}{\mathcal{Z}_{|1\rangle}} \exp\left[ -\boldsymbol{x}^\dagger \boldsymbol{M}_{2n} \boldsymbol{x} + i\eta \sum_{j=1}^{2n} (x_j + x_j^*) \right]. \tag{S17}$$

The probability distribution $p_{|1\rangle}(x_j, x_j^*)$, of the $j$-th vertex of the Gaussian distribution (S17), gives thus the solution to our original problem $\eta \langle j | \boldsymbol{M}_{2n}^{-1}(\lambda, \eta) | 1 \rangle$. This implies that we have translated a matrix inversion problem into an inference problem.

### 2. Gaussian belief propagation for sparse random matrices

With Gaussian belief propagation it is possible to infer the single-vertex marginal of a Gaussian distribution on a graph [7, 8]. Here we apply this method to the inference of $p_{|1\rangle}(x_j, x_j^*)$ [3, 9].

Before proceeding, we introduce a convenient change of variables. We define the two-dimensional vectors

$$\mathsf{x}_j = \begin{pmatrix} x_{j+n} \\ x_j \end{pmatrix} \quad j = 1, \ldots, n, \tag{S18}$$

and write the Gaussian distribution, given by Eq. (S17), in terms of the variables $\mathsf{x}_1, \ldots, \mathsf{x}_n$:

$$p_{|1\rangle}\left(\{\mathsf{x}_j, \mathsf{x}_j^\dagger\}_{j=1,\ldots,n}\right) = \frac{1}{\mathcal{Z}_1} \exp\left[ -i \sum_{k=1}^n \mathsf{x}_k^\dagger (-i\eta \mathbf{1}_2 + \boldsymbol{\lambda}) \mathsf{x}_k + i \sum_{k,j=1}^n \mathsf{x}_k^\dagger \mathsf{A}_{kj} \mathsf{x}_j + i\eta \sum_{k=1}^n \left( \mathsf{u}^\dagger \mathsf{x}_k + \mathsf{x}_k^\dagger \mathsf{u} \right) \right], \tag{S19}$$

with the two-dimensional vector $\mathsf{u} = (1\ 1)^T$. The $2 \times 2$ matrices $\boldsymbol{\lambda}$ and $\mathsf{A}_{kj}$ are given by:

$$\mathsf{A}_{kj} = \begin{pmatrix} 0 & A_{kj} \\ A_{jk}^* & 0 \end{pmatrix}, \quad \boldsymbol{\lambda} = \begin{pmatrix} 0 & \lambda \\ \lambda^* & 0 \end{pmatrix}. \tag{S20}$$

Equation (S19) is a Gaussian distribution of two-dimensional state vectors $\mathsf{x}_1, \ldots, \mathsf{x}_n$ placed on the nodes of a random graph, and interacting through the couplings $\{\mathsf{A}_{kj}\}_{k,j=1,\ldots,n}$. The single-vertex marginal $p_{|1\rangle}(\mathsf{x}_j, \mathsf{x}_j^\dagger)$ is the marginal of a Gaussian distribution, and can thus be parametrized in terms of a square matrix $\mathsf{G}_j$, and a column vector $\mathsf{H}_j$, both of dimension two:

$$p_{|1\rangle}(\mathsf{x}_j, \mathsf{x}_j^\dagger) \sim \exp\left( -i\mathsf{x}_j^\dagger (\mathsf{G}_j)^{-1} \mathsf{x}_j + i\mathsf{x}_j^\dagger \mathsf{H}_j + i(\mathsf{H}_j)^\dagger \mathsf{x}_j \right), \tag{S21}$$

where $j = 1, \ldots, n$. The approximation symbol in Eq. (S21) means that the distribution $p_{|1\rangle}(\mathsf{x}_j, \mathsf{x}_j^\dagger)$ is normalized, i.e., $\int d\mathsf{x}_j d\mathsf{x}_j^\dagger p_{|1\rangle}(\mathsf{x}_j, \mathsf{x}_j^\dagger) = 1$, with $d\mathsf{x}_j d\mathsf{x}_j^\dagger \equiv dx_j dx_j^* dx_{j+n} dx_{j+n}^*$. In terms of these new variables, all components of

$\eta \boldsymbol{M}_{2N}^{-1}(\lambda,\eta)|1\rangle$ are expressed as

$$
\begin{aligned}
\eta\langle j|\boldsymbol{M}_{2n}^{-1}|1\rangle &= -i\int \mathrm{d}\mathsf{x}_j \mathrm{d}\mathsf{x}_j^\dagger \,(\mathsf{x}_j)_1\, p_{|1\rangle}(\mathsf{x}_j, \mathsf{x}_j^\dagger) \\
&= -i\,(\mathsf{G}_j \mathsf{H}_j)_1 \,, 
\end{aligned}
\tag{S22}
$$

$$
\begin{aligned}
\eta\langle j+n|\boldsymbol{M}_{2n}^{-1}|1\rangle &= -i\int \mathrm{d}\mathsf{x}_j \mathrm{d}\mathsf{x}_j^\dagger \,(\mathsf{x}_j)_2\, p_{|1\rangle}(\mathsf{x}_j, \mathsf{x}_j^\dagger) \,,\\
&= -i\,(\mathsf{G}_j \mathsf{H}_j)_2 \,.
\end{aligned}
\tag{S23}
$$

We identify the components of the isolated eigenvectors according to Eq. (S11), and find a relation for the eigenvector components as a function of the variables $\mathsf{G}_j$ and $\mathsf{H}_j$:

$$
\langle j|l_{\mathrm{isol}}\rangle = -i \lim_{\eta\to 0^+} (\mathsf{G}_j \mathsf{H}_j)_1 \Big|_{\lambda=\lambda_{\mathrm{isol}}} \,,
\tag{S24}
$$

$$
\langle j|r_{\mathrm{isol}}\rangle = -i \lim_{\eta\to 0^+} (\mathsf{G}_j \mathsf{H}_j)_2 \Big|_{\lambda=\lambda_{\mathrm{isol}}} \,.
\tag{S25}
$$

The next step consists in deriving a relation for the variables $\mathsf{G}_j$ and $\mathsf{H}_j$, or equivalently, the marginal $p_{|1\rangle}(\mathsf{x}_j, \mathsf{x}_j^\dagger)$. We thus need to solve following inference problem

$$
p_{|1\rangle}(\mathsf{x}_j, \mathsf{x}_j^\dagger) = \int \left(\prod_{k=1(k\neq j)}^n \mathrm{d}\mathsf{x}_k \mathrm{d}\mathsf{x}_k^\dagger\right) p_{|1\rangle}\left(\{\mathsf{x}_\ell, \mathsf{x}_\ell^\dagger\}_{\ell=1,\ldots,n}\right) \,,
\tag{S26}
$$

with the distribution $p_{|1\rangle}\left(\{\mathsf{x}_\ell, \mathsf{x}_\ell^\dagger\}_{\ell=1,\ldots,n}\right)$ given by Eq. (S19). We separate in Eq. (S19) the terms coming from the $j$-th vertex from the other terms

$$
\begin{aligned}
p_{|1\rangle}\left(\{\mathsf{x}_\ell, \mathsf{x}_\ell^\dagger\}_{\ell=1,\ldots,n}\right) &\sim \exp\left[-i\mathsf{x}_j^\dagger\left(-i\eta\mathbf{1}_2 + \boldsymbol{\lambda} - \mathsf{A}_{jj}\right)\mathsf{x}_j + i\,\eta\,\mathsf{u}^\dagger \mathsf{x}_j + i\,\eta\,\mathsf{x}_j^\dagger \mathsf{u}\right] \\
&\times \exp\left[i\mathsf{x}_j^\dagger \sum_{k\in\partial_j} \mathsf{A}_{jk}\mathsf{x}_k + i\sum_{k\in\partial_j} \mathsf{x}_k^\dagger \mathsf{A}_{kj}\mathsf{x}_j\right] p_{|1\rangle}^{(j)}\left(\{\mathsf{x}_k, \mathsf{x}_k^\dagger\}_{k=1,\ldots,n;k\neq j}\right) \,,
\end{aligned}
\tag{S27}
$$

where $\partial_j$ is the set of nodes adjacent to $j$ (i.e., the indices $k$ for which either $A_{kj}\neq 0$ or $A_{jk}\neq 0$); $p_{|1\rangle}^{(j)}\left(\{\mathsf{x}_k, \mathsf{x}_k^\dagger\}_{k=1,\ldots,n;\ell\neq j}\right)$ is the distribution associated to the cavity matrix $\boldsymbol{A}_{n-1}^{(j)}$. The cavity matrix $\boldsymbol{A}_{n-1}^{(j)}$ is the $(n-1)$-dimensional submatrix of $\boldsymbol{A}_n$, which we obtain from $\boldsymbol{A}_n$ by removing the $j$-th row and the $j$-th column [4]. The approximation symbol in Eq. (S27) means again normalization. Inserting Eq. (S27) in Eq. (S26), and integrating out all variables $\mathsf{x}_k$ and $\mathsf{x}_k^\dagger$ with $k\notin\partial_j\cup\{j\}$, we get

$$
\begin{aligned}
p_{|1\rangle}(\mathsf{x}_j, \mathsf{x}_j^\dagger) &\sim \exp\left[-i\mathsf{x}_j^\dagger\left(-i\eta\mathbf{1}_2 + \boldsymbol{\lambda} - \mathsf{A}_{jj}\right)\mathsf{x}_j + i\eta\mathsf{u}^\dagger \mathsf{x}_j + i\eta\mathsf{x}_j^\dagger \mathsf{u}\right] \\
&\times \int \left(\prod_{k\in\partial_j} \mathrm{d}\mathsf{x}_k \mathrm{d}\mathsf{x}_k^\dagger\right) \exp\left[i\mathsf{x}_j^\dagger \sum_{k\in\partial_j} \mathsf{A}_{jk}\mathsf{x}_k + i\sum_{k\in\partial_j} \mathsf{x}_k^\dagger \mathsf{A}_{kj}\mathsf{x}_j\right] p_{|1\rangle}^{(j)}(\mathsf{x}_{\partial_j}, \mathsf{x}_{\partial_j}^\dagger) \,,
\end{aligned}
\tag{S28}
$$

with $p_{|1\rangle}^{(j)}(\mathsf{x}_{\partial_j}, \mathsf{x}_{\partial_j}^\dagger)$ denoting the joint distribution of the state variables belonging to the $\partial_j$. For sparse random ensembles in the limit $n\to\infty$, the vertices in the neighbourhood set $\partial_j$ form disconnected branches. This implies that the joint distribution $p_{|1\rangle}^{(j)}(\mathsf{x}_{\partial_j}, \mathsf{x}_{\partial_j}^\dagger)$ factorizes

$$
p_{|1\rangle}^{(j)}(\mathsf{x}_{\partial_j}, \mathsf{x}_{\partial_j}^\dagger) = \prod_{k\in\partial_j} p_{|1\rangle}^{(j)}(\mathsf{x}_k, \mathsf{x}_k^\dagger) \,,
\tag{S29}
$$

in the infinite size limit. The above independence property is the essence of Gaussian belief propagation, and is here a consequence of the local tree structure of $\boldsymbol{A}_n$. Substituting Eq. (S29) back in Eq. (S28) leads to

$$
\begin{aligned}
p_{|1\rangle}(\mathsf{x}_j, \mathsf{x}_j^\dagger) &\sim \exp\left[-i\mathsf{x}_j^\dagger\left(-i\eta\mathbf{1}_2 + \boldsymbol{\lambda} - \mathsf{A}_{jj}\right)\mathsf{x}_j + i\eta\mathsf{u}^\dagger \mathsf{x}_j + i\eta\mathsf{x}_j^\dagger \mathsf{u}\right] \\
&\times \prod_{k\in\partial_j}\left[\int \mathrm{d}\mathsf{x}_k \mathrm{d}\mathsf{x}_k^\dagger \exp\left(i\mathsf{x}_j^\dagger \mathsf{A}_{jk}\mathsf{x}_k + i\mathsf{x}_k^\dagger \mathsf{A}_{kj}\mathsf{x}_j\right) p_{|1\rangle}^{(j)}(\mathsf{x}_k, \mathsf{x}_k^\dagger)\right] \,.
\end{aligned}
\tag{S30}
$$

In order to proceed further, we use the following parametrization of the marginals $p_{|1\rangle}^{(j)}(\mathsf{x}_k, \mathsf{x}_k^\dagger)$

$$p_{|1\rangle}^{(j)}(\mathsf{x}_k, \mathsf{x}_k^\dagger) \sim \exp\left(-i\mathsf{x}_k^\dagger \left(\mathsf{G}_k^{(j)}\right)^{-1} \mathsf{x}_k + i\mathsf{x}_k^\dagger \mathsf{H}_k^{(j)} + i(\mathsf{H}_k^{(j)})^\dagger \mathsf{x}_k\right), \tag{S31}$$

with $\mathsf{G}_k^{(j)}$ and $\mathsf{H}_k^{(j)}$ matrices and column vectors of dimension 2; the variables $\mathsf{G}_k^{(j)}$ and $\mathsf{H}_k^{(j)}$ for the matrix $\boldsymbol{A}_{n-1}^{(j)}$ are equivalent to the variables $\mathsf{G}_k$ and $\mathsf{H}_k$ for the matrix $\boldsymbol{A}_n$. Inserting the above relation in Eq. (S30) and integrating out the state variables $\{\mathsf{x}_k, \mathsf{x}_k^\dagger\}_{k \in \partial_j}$, we obtain

$$p_{|1\rangle}(\mathsf{x}_j, \mathsf{x}_j^\dagger) \sim \exp\left[-i\mathsf{x}_j^\dagger \left(-i\eta \mathbf{1}_2 + \boldsymbol{\lambda} - \mathsf{A}_{jj} - \sum_{k\in\partial_j} \mathsf{A}_{jk} \mathsf{G}_k^{(j)} \mathsf{A}_{kj}\right) \mathsf{x}_j\right]$$
$$\times \exp\left[i\mathsf{x}_j^\dagger \left(\eta\,\mathsf{u} + \sum_{k\in\partial_j} \mathsf{A}_{jk} \mathsf{G}_k^{(j)} \mathsf{H}_k^{(j)}\right) + i \left(\eta\,\mathsf{u} + \sum_{k\in\partial_j} \mathsf{A}_{jk} \mathsf{G}_k^{(j)} \mathsf{H}_k^{(j)}\right)^\dagger \mathsf{x}_j\right]. \tag{S32}$$

Finally, we substitute Eq. (S32) in Eqs. (S22) and (S23), and find a set of equations for $\mathsf{G}_j$ and $\mathsf{H}_j$:

$$\mathsf{G}_j = \left(-i\eta\,\mathbf{1}_2 + \boldsymbol{\lambda} - \mathsf{A}_{jj} - \sum_{k\in\partial_j} \mathsf{A}_{jk} \mathsf{G}_k^{(j)} \mathsf{A}_{kj}\right)^{-1}, \tag{S33}$$

$$\mathsf{H}_j = \eta\,\mathbf{1}_2 + \sum_{k\in\partial_j} \mathsf{A}_{jk} \mathsf{G}_k^{(j)} \mathsf{H}_k^{(j)}. \tag{S34}$$

The random variables $\mathsf{G}_j$ and $\mathsf{H}_j$ depend on $\{\mathsf{G}_k^{(j)}\}_{k\in\partial_j}$ and $\{\mathsf{H}_k^{(j)}\}_{k\in\partial_j}$, both defined on a the cavity matrix $\boldsymbol{A}_{n-1}^{(j)}$. At this level of our approach, these variables remain undetermined. and we need a closed set of equations for the single site marginals $p_{|1\rangle}^{(j)}(\mathsf{x}_k, \mathsf{x}_k^\dagger)$.

In order to derive a set of equations in the random variables $\mathsf{G}_j^{(\ell)}$ and $\mathsf{H}_k^{(\ell)}$, with $\ell \in \partial_j$, we repeat the same procedure as described above, but we apply it now to the cavity matrix $\boldsymbol{A}_{n-1}^{(\ell)}$. In other words, we are developing a recursive procedure. We have now the following inference problem to solve

$$p_{|1\rangle}^{(\ell)}(\mathsf{x}_j, \mathsf{x}_j^\dagger) = \int \left(\prod_{k=1(k\neq j, k\neq\ell)}^{2n} d\mathsf{x}_k d\mathsf{x}_k^\dagger\right) p_{|1\rangle}^{(\ell)}(\{\mathsf{x}_k, \mathsf{x}_k^\dagger\}), \;\; \ell \in \partial_j. \tag{S35}$$

The marginal $p_{|1\rangle}^{(\ell)}(\mathsf{x}_j, \mathsf{x}_j^\dagger)$ on the cavity graph is given by an expression similar to the right hand side of Eq. (S30), but in this case the product runs over $\partial_j \setminus \ell$, since $\ell \in \partial_j$. After following the aforementioned program, we find that the random variables $\mathsf{G}_j^{(\ell)}$ and $\mathsf{H}_j^{(\ell)}$ solve the following equations:

$$\mathsf{G}_j^{(\ell)} = \left(-i\eta\,\mathbf{1}_2 + \boldsymbol{\lambda} - \mathsf{A}_{jj} - \sum_{k\in\partial_j\setminus\{\ell\}} \mathsf{A}_{jk} \mathsf{G}_k^{(j,\ell)} \mathsf{A}_{kj}\right)^{-1}, \tag{S36}$$

$$\mathsf{H}_j^{(\ell)} = \eta\,\mathbf{1}_2 + \sum_{k\in\partial_j\setminus\{\ell\}} \mathsf{A}_{jk} \mathsf{G}_k^{(j,\ell)} \mathsf{H}_k^{(j,\ell)}. \tag{S37}$$

Equations (S36) and (S37) are equivalent to the Eqs. (S33) and (S34), with the difference that Eqs. (S36)-(S37) apply to the matrix $\boldsymbol{A}_n$, whereas the Eqs. (S33)-(S34) apply to the cavity matrix $\boldsymbol{A}_n^{(\ell)}$. We see that the random variables $\mathsf{G}_j^{(\ell)}$ and $\mathsf{H}_j^{(\ell)}$ are now expressed in terms of random variables $\mathsf{G}_k^{(j,\ell)}$ and $\mathsf{H}_k^{(j,\ell)}$. We need to repeat the recursion procedure to find expressions for the random variables $\mathsf{G}_k^{(j,\ell)}$ and $\mathsf{H}_k^{(j,\ell)}$.

In order to close the recursion, we set $\mathsf{G}^{(j,\ell)} = \mathsf{G}^{(j)}$ and $\mathsf{H}^{(j,\ell)} = \mathsf{H}^{(j)}$. These identities are valid for sparse ensembles

in the limit $n \to \infty$, because they are local tree like. The Eqs. (S36) and (S37) become:

$$\mathsf{G}_j^{(\ell)} = \left(-i\eta\, \mathbf{1}_2 + \boldsymbol{\lambda} - \mathsf{A}_{jj} - \sum_{k \in \partial_j \setminus \{\ell\}} \mathsf{A}_{jk} \mathsf{G}_k^{(j)} \mathsf{A}_{kj}\right)^{-1}, \tag{S38}$$

$$\mathsf{H}_j^{(\ell)} = \eta\, \mathbf{1}_2 + \sum_{k \in \partial_j \setminus \{\ell\}} \mathsf{A}_{jk} \mathsf{G}_k^{(j)} \mathsf{H}_k^{(j)}. \tag{S39}$$

Equations (S33), (S34), (S38) and (S39) form a closed set of equations that determine the spectral properties of sparse non-Hermitian matrices. The limit $\eta \to 0^+$ is implicit in these equations.

The outlier $\lambda_{\text{isol}}$ is given by values $\lambda$ for which the Eqs. (S34) and (S39) admit a non-trivial solution, i.e., a solution for which $\mathsf{H}_j \neq 0$ ($\forall j$) or $\mathsf{H}_j \neq \pm\infty$ ($\forall j$). This non-trivial solution determines then the eigenvectors associated to the outlier $\lambda_{\text{isol}}$ via Eqs. (S24)-(S25). Apart from the outlier, the Eqs. (S33) and (S38) determine also the spectral density of the ensemble via $\rho(\lambda) = (n\pi)^{-1} \lim_{\eta \to 0^+} \partial^* \sum_{j=1}^n (\mathsf{G}_j)_{21}$ [3, 4].

Note that our formalism unifies a theory for the outliers of sparse non-Hermitian matrices with the theory for the spectral density function $\rho(\lambda)$ of Refs. [3, 4]. In the following section, we show that the resultant theory can be simplified.

### C. Simplification for $\lambda \notin \Omega$

Outliers lie, by definition, outside the support $\Omega$ of the spectral density. Since the resolvent is well defined outside the support of the spectral density, the Eqs. (S33), (S34), (S38) and (S39) simplify for values $\lambda \notin \Omega$. The Hermitization procedure provides advantage in our calculations, which deal with normal matrices. But, conceptually there seems no need to regularize the ensemble $\boldsymbol{A}_n$ outside the support of the spectral density.

Here we show how that, indeed, the final Eqs. (S33), (S34), (S38) and (S39) on the regularized ensemble $\boldsymbol{M}_{2n}$, can be simplified into equations of random variables defined on the original ensemble $\boldsymbol{A}_n$. The final result of this procedure are the Eqs. (5)-(10) in the main text.

The Eqs. (S33) and (S38) admit the solution

$$\mathsf{G}_j = \begin{pmatrix} 0 & -g_j^* \\ -g_j & 0 \end{pmatrix}, \quad \mathsf{G}_j^{(\ell)} = \begin{pmatrix} 0 & -\left(g_j^{(\ell)}\right)^* \\ -g_j^{(\ell)} & 0 \end{pmatrix}, \tag{S40}$$

which is stable for all values $\lambda \notin \Omega$. The random variables $g_j$ are the diagonal elements of the resolvent $(\boldsymbol{A}_n - \lambda \mathbf{1}_n)^{-1}$, i.e., $g_j = \left[(\boldsymbol{A}_n - \lambda \mathbf{1}_n)^{-1}\right]_{jj}$. Analogously, the random variables $g_j^{(\ell)}$ are the diagonal elements of the resolvent of the cavity matrix $\boldsymbol{A}_{n-1}^{(\ell)}$, i.e., $g_j^{(\ell)} = \left[(\boldsymbol{A}_{n-1}^{(\ell)} - \lambda \mathbf{1}_n)^{-1}\right]_{jj}$.

Since here we are interested in the spectral properties of the outlier, which satisfies $\lambda_{\text{isol}} \notin \Omega$, we use the simplified solution, given by Eq. (S40), in the Eqs. (S33) and (S38). We find that $g_j$ and $g_j^{(\ell)}$ solve the equations

$$g_j = \frac{1}{-\lambda + i\eta + A_{jj} - \sum_{k \in \partial_i} A_{jk} g_k^{(j)} A_{kj}}, \tag{S41}$$

$$g_j^{(\ell)} = \frac{1}{-\lambda + i\eta + A_{jj} - \sum_{k \in \partial_i \setminus \{\ell\}} A_{jk} g_k^{(j)} A_{kj}}, \tag{S42}$$

for $\lambda \notin \Omega$, and for which the limit $\eta \to 0^+$ is implicit. Note that for real and symmetric matrices the resolvent relations (S118)-(S116) are identical to the resolvent equations in the Refs. [10, 12, 12–14].

The relations for the random variables $\mathsf{H}_j$ and $\mathsf{H}_j^{(\ell)}$, given by (S34) and (S39), simplify as well when we use solution (S40). Let us first introduce some novel notation. We write for the elements of the right- and left-eigenvectors of $\lambda_{\text{isol}}$, respectively, $r_j = \langle j | r_{\text{isol}} \rangle$ and $l_j = \langle j | l_{\text{isol}} \rangle$. From Eqs. (S24)-(S25) it follows that

$$l_j = -i \lim_{\eta \to 0^+} (\mathsf{G}_j \mathsf{H}_j)_1 \Big|_{\lambda = \lambda_{\text{isol}}}, \tag{S43}$$

$$r_j = -i \lim_{\eta \to 0^+} (\mathsf{G}_j \mathsf{H}_j)_2 \Big|_{\lambda = \lambda_{\text{isol}}}. \tag{S44}$$

When we apply the same arguments to the cavity matrix $\boldsymbol{A}_{n-1}^{(\ell)}$, we find

$$l_j^{(\ell)} = -i \lim_{\eta \to 0^+} \left(\mathsf{G}_j^{(\ell)} \mathsf{H}_j^{(\ell)}\right)_1 \bigg|_{\lambda = \lambda_{\text{isol}}}, \tag{S45}$$

$$r_j^{(\ell)} = -i \lim_{\eta \to 0^+} \left(\mathsf{G}_j^{(\ell)} \mathsf{H}_j^{(\ell)}\right)_2 \bigg|_{\lambda = \lambda_{\text{isol}}}, \tag{S46}$$

for the eigenvector elements $r_j^{(\ell)} = \langle j | r_{\text{isol}}^{(\ell)} \rangle$ and $l_j^{(\ell)} = \langle j | l_{\text{isol}}^{(\ell)} \rangle$ of the eigenvectors associated to the eigenvalue $\lambda_{\text{isol}}$ of the cavity matrix $\boldsymbol{A}_{n-1}^{(\ell)}$. After substitution of Eq. (S40) in the Eqs. (S34) and (S39), we find that the random variables $r_j$, $l_j$, $r_j^{(\ell)}$ and $l_j^{(\ell)}$ solve the equations:

$$r_j = -g_j \sum_{k \in \partial_j} A_{jk} r_k^{(j)}, \tag{S47}$$

$$l_j^* = -g_j \sum_{k \in \partial_j} \left(l_k^{(j)}\right)^* A_{kj}, \tag{S48}$$

$$r_j^{(\ell)} = -g_j^{(\ell)} \sum_{k \in \partial_j \setminus \{\ell\}} A_{jk} r_k^{(j)}, \tag{S49}$$

$$\left(l_j^{(\ell)}\right)^* = -g_j^{(\ell)} \sum_{k \in \partial_j \setminus \{\ell\}} \left(l_k^{(j)}\right)^* A_{kj}. \tag{S50}$$

We have thus derived the Eqs. (5)-(10) in the main text.

## S5. THEORY FOR ORIENTED RANDOM MATRICES

We demonstrate our theory on oriented random matrices. Oriented matrices are matrices for which all links between two vertices are directed, i.e., for which $C_{jk} C_{kj} = 0$ for all values $j \neq k$.

In subsection A, we apply the relations (5)-(10) to oriented matrix ensembles. In subsection B, we derive a closed set of equations for the average of the distribution of eigenvector elements $r_j$ and $l_j$ associated to the outlier $\lambda_{\text{isol}}$, and we discuss how to solve this equation numerically with the population dynamics algorithm. In subsection C, we present a closed set of equations for the first moments the average distribution of eigenvector elements. In the last subsection D, we derive an exact equation for the boundary of the support of the spectral density function.

### A. Generic equations for an oriented random matrix

For an oriented random matrix, the Eqs. (S118)-(S116) simplify considerably. We find the following explicit expression for the resolvent elements

$$g_j = g_j^{(\ell)} = \frac{1}{-\lambda + i\eta + A_{jj}}. \tag{S51}$$

We use the explicit expression for the resolvent elements, given by Eq. S51, in the Eqs. (S123)-(S50) for the right- and left-eigenvector elements:

$$r_j = r_j^{(\ell)}, \quad \ell \in \partial_j^{\text{in}}, \tag{S52}$$

$$l_j = l_j^{(\ell)}, \quad \ell \in \partial_k^{\text{out}}, \tag{S53}$$

$$r_j^{(\ell)} = \frac{1}{(\lambda - J_{jj})} \sum_{k \in \partial_j^{\text{out}}} J_{jk} r_k^{(j)}, \tag{S54}$$

$$\left(l_j^{(\ell)}\right)^* = \frac{1}{(\lambda - J_{jj})} \sum_{k \in \partial_j^{\text{in}}} J_{kj} \left(l_k^{(j)}\right)^*. \tag{S55}$$

Here, the in-neighbourhood $\partial_j^{\text{in}}$ of the $j$-th vertex is the set of vertices $k \neq j$ with $C_{kj} \neq 0$, and the out-neighbourhood $\partial_j^{\text{out}}$ of the $j$-th vertex is the set of vertices $k \neq j$ with $C_{jk} \neq 0$.

The Eqs. (S54) express the random variables $r_j$ and $r_j^{(\ell)}$ in terms of local random variables that are statistically independent for $n \to \infty$. Indeed, in this limit the variables $\left\{r_k^{(j)}\right\}_{k \in \partial_j^{\text{out}}}$ are mutually independent, and they are also independent of the couplings $\{J_{jk}\}_{k \in \partial_j^{\text{out}}}$. A similar statement holds for the Eqs. (S55). The statistical independence of the random variables on the right hand side of Eqs. (S54) and (S55) is important, since it allows us to derive a closed and self-consistent equation for the ensemble-averaged distribution of eigenvector elements; this statistical independency allows us also to derive a closed set of equations in the first moments of the average distribution of eigenvector elements. Oriented random matrices are thus exactly solvable, in the sense that we can derive exact analytical expressions for several spectral quantities, such as the eigenvalue outlier and the first moments of the associated eigenvectors. Note that the fact that oriented matrices are exactly solvable seems to be related to the fact that the dynamics of spin models on oriented graphs is also exactly solvable [15, 16]. In the following subsections we derive a closed set of equations for the ensemble averaged distribution of eigenvector elements.

### B. Population dynamics algorithm

We derive a closed set of equations in the distribution of eigenvector elements $p_{\text{R,L}}(r,l)$ associated to the outlier $\lambda_{\text{isol}}$. We define the distribution of eigenvector elements as

$$p_{\text{R,L}}(r,l) = \left\langle n^{-1} \sum_{j=1}^n \delta(r - r_j)\delta(l - l_j) \right\rangle. \tag{S56}$$

We take the average of Eqs. (S52)-(S55) over the ensemble $p(\boldsymbol{A}_n)$ and find that $p_{\text{R,L}}(r,l)$ solves the following equation

$$p_{\text{R,L}}(r,l) = \sum_{K^{\text{in}}=0}^{\infty} \sum_{K^{\text{out}}=0}^{\infty} p_{\text{deg}}(K^{\text{in}}, K^{\text{out}})$$

$$\int dD \prod_{k=1}^{K^{\text{out}}} dJ_k \prod_{k=1}^{K^{\text{in}}} dJ'_k \; p\left(D, J_1, J_2, \ldots, J_{K^{\text{out}}}, J'_1, J'_2, \ldots, J'_{K^{\text{in}}}\right)$$

$$\int \prod_{k=1}^{K^{\text{out}}} dr_k \prod_{k=1}^{K^{\text{in}}} dl_k \prod_{k=1}^{K^{\text{out}}} p_{\text{R}}(r_k) \prod_{k=1}^{K^{\text{in}}} p_{\text{L}}(l_k)$$

$$\delta\left(r - \frac{1}{(\lambda - D)} \sum_{k=1}^{K^{\text{out}}} J_k r_k\right) \delta\left(l - \frac{1}{(\lambda^* - D^*)} \sum_{k=1}^{K^{\text{in}}} (J'_k)^* l_k\right). \tag{S57}$$

Here $p_{\text{deg}}(K^{\text{in}}, K^{\text{out}})$ is the distribution of in- and out-degrees. Recall that the random variables $K_j^{\text{in}}$ and $K_j^{\text{out}}$ are, respectively, the indegree and the outdegree of the $j$-th vertex. They are defined as the size of the sets $\partial_j^{\text{in}}$ and $\partial_j^{\text{out}}$, respectively. The degrees $(K_j^{\text{in}}, K_j^{\text{out}})$ are thus random variables taken from the distribution $p_{\text{deg}}(K^{\text{in}}, K^{\text{out}})$.

We have also introduced a new notation in order to distinguish clearly between diagonal and off-diagonal matrix elements: $D$ represents an arbitrary diagonal element, whereas the random variables $\{J_j\}$ and $\{J'_j\}$ denote off-diagonal elements. In the language of graph theory, $J_1, \ldots, J_{K^{\text{out}}}$ ($J'_1, \ldots, J'_{K^{\text{in}}}$) are the weights of the outgoing (incoming) links of a certain node. The function $p\left(D, J_1, J_2, \ldots, J_{K^{\text{out}}}, J'_1, J'_2, \ldots, J'_{K^{\text{in}}}\right)$ is the joint distribution of non-zero matrix elements associated to one vertex. We have also introduced the distributions $p_{\text{L}}(l) = \int dr \; p_{\text{R,L}}(r,l)$, and $p_{\text{R}}(r) = \int dl \; p_{\text{R,L}}(r,l)$.

We solve the Eq. (S57) through a Monte Carlo method, which is often called the population dynamics algorithm [10, 12, 13, 17]. This algorithm sovles the Eq. (S57) by representing the unknown distribution $p_{\text{R,L}}(r,l)$ as a population of $M$ samples $\{(r_j, l_j)\}_{j=1,\ldots,M}$. We evolve this population randomly and sequentially according to a dynamical process. The stationary state of this process corresponds to the solution to the Eq. (S57). The algorithm consists in repeating the following steps sequentially. We first sample two random degrees $K^{\text{in}}$ and $K^{\text{out}}$ from the distribution $p_{\text{deg}}(K^{\text{in}}, K^{\text{out}})$; we then sample a set of elements $D$, $\{J_k\}_{k=1,\ldots,K^{\text{out}}}$ and $\{J'_k\}_{k=1,\ldots,K^{\text{in}}}$ from the distribution $p\left(D, J_1, J_2, \ldots, J_{K^{\text{out}}}, J'_1, J'_2, \ldots, J'_{K^{\text{in}}}\right)$; we further sample $K^{\text{in}}$ elements $l_k$ and $K^{\text{out}}$ elements $r_k$ from the population of $M$ samples. After having sampled all these random variables, we replace one randomly chosen sample $(r_j, l_j)$ in

the population by the value

$$(r_j, l_j) = \left( \frac{1}{(\lambda - D)} \sum_{k=1}^{K^{\text{out}}} J_k r_k, \frac{1}{(\lambda^* - D^*)} \sum_{k=1}^{K^{\text{in}}} (J'_k)^* l_k \right). \tag{S58}$$

We repeat the above steps sequentially until the population has reached a stationary state. In the stationary state the population consist of $M$ random samples taken from the distribution $p_{\text{R,L}}(r, l)$ that solves the Eq. (S57).

### C. Ensemble averages for the moments of the eigenvector distribution

We derive a closed set of equations in the first moments of the eigenvector distribution $p_{\text{R,L}}(r, l)$. We write for the average of a function $f(r, l)$: $\langle f(r, l) \rangle = \int dr dl f(r, l) p_{\text{R,L}}(r, l)$. We take the ensemble average of the Eqs. (S54)-(S55), and find the following equations for the first and second moments of the distribution of eigenvector elements Eq. (S56):

$$\langle r \rangle = \langle r \rangle \left\langle \frac{\sum_{j=1}^{K^{\text{out}}} J_j}{\lambda - D} \right\rangle, \tag{S59}$$

$$\langle l \rangle = \langle l \rangle \left\langle \frac{\sum_{j=1}^{K^{\text{in}}} (J'_j)^*}{\lambda^* - D^*} \right\rangle, \tag{S60}$$

$$\langle |r|^2 \rangle = \langle |r|^2 \rangle \left\langle \frac{\sum_{j=1}^{K^{\text{out}}} |J_j|^2}{|\lambda - D|^2} \right\rangle + |\langle r \rangle|^2 \left\langle \frac{\sum_{k \neq j=1}^{K^{\text{out}}} J_k J_j^*}{|\lambda - D|^2} \right\rangle, \tag{S61}$$

$$\langle |l|^2 \rangle = \langle |l|^2 \rangle \left\langle \frac{\sum_{j=1}^{K^{\text{in}}} |J'_j|^2}{|\lambda - D|^2} \right\rangle + |\langle l \rangle|^2 \left\langle \frac{\sum_{k \neq j=1}^{K^{\text{in}}} J'_k (J'_j)^*}{|\lambda - D|^2} \right\rangle, \tag{S62}$$

$$\langle r^2 \rangle = \langle r^2 \rangle \left\langle \frac{\sum_{j=1}^{K^{\text{out}}} J_j^2}{(\lambda - D)^2} \right\rangle + \langle r \rangle^2 \left\langle \frac{\sum_{k \neq j=1}^{K^{\text{out}}} J_k J_j}{(\lambda - D)^2} \right\rangle, \tag{S63}$$

$$\langle l^2 \rangle = \langle l^2 \rangle \left\langle \frac{\sum_{j=1}^{K^{\text{in}}} ((J'_j)^*)^2}{(\lambda^* - D^*)^2} \right\rangle + \langle l \rangle^2 \left\langle \frac{\sum_{k \neq j=1}^{K^{\text{in}}} (J'_k)^* (J'_j)^*}{(\lambda^* - D^*)^2} \right\rangle. \tag{S64}$$

The symbol $\langle \ldots \rangle$ stands for the average over the joint distribution of the degrees $(K^{\text{in}}, K^{\text{out}})$, the off-diagonal elements $\{J_j\}$ and $\{J'_j\}$, and the diagonal elements $D$.

The solutions of the above set of equations describe the first two eigenvector moments of a broad class of oriented random matrices, including the case where the matrix elements are correlated. In Section S6 we solve the above equations analytically for ensembles of adjacency matrices with statistically independent off-diagonal and diagonal elements, and in Section S7 we solve the above equations for the ensemble of Laplacian random matrices characterizing a continuous time Markov process.

### D. Boundary of the support of the spectral density

We derive a set of equations for the boundary $\partial \Omega$ of the support of the spectral density of oriented matrices. We write $\lambda_b$ for values of $\lambda$ located at the boundary of the support, i.e., $\lambda_b \in \partial \Omega$. The approach here is based on a stability analysis of the solution (S40) to the Eqs. (S33) and (S38). The solution, given by Eq. (S40), is stable for values $\lambda$ outside the support of the spectral density function, i.e., $\lambda \notin \Omega$. If however $\lambda \in \Omega$, then the solution, given by Eq. (S40), is unstable under small perturbations. An expression for the boundary of the support $\partial \Omega$ follows therefore from a linear stability analysis [4].

When we apply a linear stability analysis of the solution (S40) to the Eqs. (S33) and (S36) we find the stability

criteria

$$\left\langle \frac{\sum_{k\in\partial_j^{\text{out}}} |A_{jk}|^2}{|\lambda - A_{jj}|^2} \right\rangle_{\boldsymbol{A}_n} \leq 1, \tag{S65}$$

$$\left\langle \frac{\sum_{k\in\partial_j^{\text{in}}} |A_{kj}|^2}{|\lambda - A_{jj}|^2} \right\rangle_{\boldsymbol{A}_n} \leq 1. \tag{S66}$$

If both of the above two stability conditions hold, the solution for the resolvent, given by Eq. (S40), is stable. The support $\Omega$ is thus the set of values $\lambda \in \mathbb{C}$ for which

$$\left\langle \frac{\sum_{k\in\partial_j^{\text{out}}} |A_{jk}|^2}{|\lambda - A_{jj}|^2} \right\rangle_{\boldsymbol{A}_n} > 1, \quad \text{or,} \quad \left\langle \frac{\sum_{k\in\partial_j^{\text{in}}} |A_{kj}|^2}{|\lambda - A_{jj}|^2} \right\rangle_{\boldsymbol{A}_n} > 1. \tag{S67}$$

If additionally $P(\boldsymbol{A}_n) = P(\boldsymbol{A}_n^\dagger)$, then Eqs. (S65) and (S66) are identical. The eigenvalues $\lambda_{\text{b}}$ at the boundary of the support solve then

$$1 = \left\langle \frac{\sum_{k\in\partial_j^{\text{out}}} |A_{jk}|^2}{|\lambda_{\text{b}} - A_{jj}|^2} \right\rangle_{\boldsymbol{A}_n}, \tag{S68}$$

and the support $\Omega$ is given by the values of $\lambda \in \mathbb{C}$ for which:

$$\left\langle \frac{\sum_{k\in\partial_j^{\text{out}}} |A_{jk}|^2}{|\lambda - A_{jj}|^2} \right\rangle_{\boldsymbol{A}_n} > 1. \tag{S69}$$

We now show how to derive the Eqs. (S65)-(S66). We perturb the solution for the resolvent elements of oriented matrices, given by the Eqs. (S40) and (S51), as follows:

$$\mathsf{G}_j^{(\ell)} = \begin{pmatrix} \epsilon_{j,1}^{(\ell)} & \lambda - A_{jj} \\ \lambda^* - A_{jj}^* & \epsilon_{j,2}^{(\ell)} \end{pmatrix}, \tag{S70}$$

with variables $|\epsilon_{i,1}^{(j)}| \ll 1$ and $|\epsilon_{i,2}^{(j)}| \ll 1$ corresponding to a small perturbation. We substitute the expression (S70) in the resolvent Eqs. (S38) and get for $\epsilon_{j,1}^{(\ell)}, \epsilon_{j,2}^{(\ell)} \to 0$:

$$\mathsf{G}_j^{(\ell)} = \left(|\lambda - A_{jj}|^2\right)^{-1} \begin{pmatrix} \sum_{k\in\partial_j\setminus\{\ell\}} \epsilon_{k,1}^{(j)} |A_{kj}|^2 & \lambda - A_{jj} \\ \lambda^* - A_{jj}^* & \sum_{k\in\partial_j\setminus\{\ell\}} \epsilon_{k,2}^{(j)} |A_{jk}|^2 \end{pmatrix} + \mathcal{O}(\epsilon^2). \tag{S71}$$

We take now the average over the ensemble $p(\boldsymbol{A}_n)$. For $\ell \in \partial_j^{\text{in}}$ we find from Eqs. (S71)

$$\begin{pmatrix} \langle \epsilon_1 \rangle_- & \langle g \rangle_- \\ \langle g^* \rangle_- & \langle \epsilon_2 \rangle_- \end{pmatrix} = \begin{pmatrix} \langle \epsilon_1 \rangle_+ \left\langle \frac{\sum_{k\in\partial_j^{\text{in}}\setminus\{\ell\}} |A_{kj}|^2}{|\lambda - A_{jj}|^2} \right\rangle & \langle (\lambda^* - A_{jj}^*)^{-1} \rangle \\ \langle (\lambda - A_{jj})^{-1} \rangle & \langle \epsilon_2 \rangle_- \left\langle \frac{\sum_{k\in\partial_j^{\text{out}}} |A_{jk}|^2}{|\lambda - A_{jj}|^2} \right\rangle \end{pmatrix}, \tag{S72}$$

and for $\ell \in \partial_j^{\text{out}}$ we find from Eqs. (S71)

$$\begin{pmatrix} \langle \epsilon_1 \rangle_+ & \langle g \rangle_+ \\ \langle g^* \rangle_+ & \langle \epsilon_2 \rangle_+ \end{pmatrix} = \begin{pmatrix} \langle \epsilon_1 \rangle_+ \left\langle \frac{\sum_{k\in\partial_j^{\text{in}}} |A_{kj}|^2}{|\lambda - A_{jj}|^2} \right\rangle & \langle (\lambda^* - A_{jj}^*)^{-1} \rangle \\ \langle (\lambda - A_{jj})^{-1} \rangle & \langle \epsilon_2 \rangle_- \left\langle \frac{\sum_{k\in\partial_j^{\text{out}}\setminus\{\ell\}} |A_{jk}|^2}{|\lambda - A_{jj}|^2} \right\rangle \end{pmatrix}. \tag{S73}$$

The Eqs. (S72)-(S73) give the stability criteria (S65) and (S66).

## S6. ORIENTED ADJACENCY MATRICES

We solve the Eqs. (S59)-(S64) analytically for ensembles of adjacency matrices with statistically independent off-diagonal and diagonal elements. We consider sparse non-Hermitian matrix ensembles for which the off-diagonal elements $A_{jk}$ are statistically independent from the diagonal matrix elements $A_{jj}$. In other words:

$$p(D, J_1, J_2, \ldots, J_{K^{\text{out}}}, J'_1, J'_2, \ldots, J'_{K^{\text{in}}}) = p_D(D) \, p(J_1, J_2, \ldots, J_{K^{\text{out}}}, J'_1, J'_2, \ldots, J'_{K^{\text{in}}}) \ . \tag{S74}$$

Additionally, we consider that the off-diagonal elements are statistical independent, such that

$$p(D, J_1, J_2, \ldots, J_{K^{\text{out}}}, J'_1, J'_2, \ldots, J'_{K^{\text{in}}}) = p_D(D) \prod_{j=1}^{K^{\text{out}}} p_J(J_j) \prod_{j=1}^{K^{\text{in}}} p_J(J'_j) \ . \tag{S75}$$

We call such random-matrix ensembles, oriented adjacency matrices. Oriented adjacency matrices obey the symmetry property $P(\boldsymbol{A}_n) = P(\boldsymbol{A}_n^\dagger)$, i.e., the distribution of $\boldsymbol{A}_n$ is invariant with respect to conjugate transposition. As a consequence, the distributions of right and left eigenvector elements, $p_R(r)$ and $p_L(l)$, are the same. We choose the following normalization for the eigenvectors

$$\langle |r|^2 \rangle = \langle |l|^2 \rangle = 1 \, , \quad \langle r \rangle \in \mathbb{R} \, , \quad \langle l \rangle \in \mathbb{R}. \tag{S76}$$

For oriented adjacency matrices the Eqs. (S59)-(S64) become (given the normalization (S76))

$$\langle r \rangle = c \langle r \rangle \langle J \rangle_J \left\langle \frac{1}{\lambda - D} \right\rangle_D, \tag{S77}$$

$$\langle l \rangle = c \langle l \rangle \langle J \rangle_J \left\langle \frac{1}{\lambda - D} \right\rangle_D, \tag{S78}$$

$$1 = c \langle |J|^2 \rangle_J \left\langle \frac{1}{|\lambda - D|^2} \right\rangle_D + \langle r \rangle^2 \left( \langle (K^{\text{out}})^2 \rangle - c \right) |\langle J \rangle_J|^2 \left\langle \frac{1}{|\lambda - D|^2} \right\rangle_D, \tag{S79}$$

$$1 = c \langle |J|^2 \rangle_J \left\langle \frac{1}{|\lambda - D|^2} \right\rangle_D + \langle l \rangle^2 \left( \langle (K^{\text{in}})^2 \rangle - c \right) |\langle J \rangle_J|^2 \left\langle \frac{1}{|\lambda - D|^2} \right\rangle_D, \tag{S80}$$

$$\langle r^2 \rangle = c \langle r^2 \rangle \langle J^2 \rangle_J \left\langle \frac{1}{(\lambda - D)^2} \right\rangle_D + \langle r \rangle^2 \left( \langle (K^{\text{out}})^2 \rangle - c \right) \langle J \rangle_J^2 \left\langle \frac{1}{(\lambda - D)^2} \right\rangle_D, \tag{S81}$$

$$\langle l^2 \rangle = c \langle l^2 \rangle \langle (J^*)^2 \rangle_J \left\langle \frac{1}{(\lambda^* - D^*)^2} \right\rangle_D + \langle l \rangle^2 \left( \langle (K^{\text{in}})^2 \rangle - c \right) \langle J^* \rangle_J^2 \left\langle \frac{1}{(\lambda^* - D^*)^2} \right\rangle_D. \tag{S82}$$

Here $c = \langle K^{\text{in}} \rangle = \langle K^{\text{out}} \rangle$ is the average value of the indegree and the outdegree. The symbols $\langle \ldots \rangle_D$ and $\langle \ldots \rangle_J$ represent, respectively, the average over the diagonal and the off-diagonal matrix elements, with distributions $p_D$ and $p_J$, respectively.

The outliers of the random matrix ensemble are given by the values of $\lambda \notin \Omega$ for which Eqs. (S77)-(S82) have a non-trivial fixed-point solution, i.e., the left and right eigenvector moments converge to well-defined limits under iteration of Eqs. (S77-S82). We find two solutions to the Eqs. (S77-S82). One solution corresponds to the outlier $\lambda_{\text{isol}}$ and a second solution $\lambda_{\text{b}}$ applies to the boundary of the support, i.e., $\lambda_{\text{b}} \in \partial\Omega$.

### A. The outlier eigenpair

The outlier is given by the value $\lambda = \lambda_{\text{isol}}$ for which the Eqs. (S77)-(S82) admit a nontrivial fixed-point solution with $\langle r \rangle \neq 0$ and $\langle l \rangle \neq 0$. From Eqs. (S77) and (S78), it follows that $\lambda_{\text{isol}}$ solves:

$$\left\langle \frac{1}{\lambda_{\text{isol}} - D} \right\rangle_D = \frac{1}{c \langle J \rangle_J}. \tag{S83}$$

Substitution of the $\lambda_{\text{isol}}$ solution, given by Eq. (S83), in Eqs. (S79) and (S80) leads to the following expressions for the first moments $\langle r \rangle$ and $\langle l \rangle$ of the eigenvector associated to $\lambda_{\text{isol}}$:

$$\langle r \rangle^2 = \frac{1}{(\langle (K^{\text{out}})^2 \rangle - c)|\langle J \rangle_J|^2} \left[ \left\langle \frac{1}{|\lambda_{\text{isol}} - D|^2} \right\rangle_D^{-1} - c \left\langle |J|^2 \right\rangle_J \right], \tag{S84}$$

$$\langle l \rangle^2 = \frac{1}{(\langle (K^{\text{in}})^2 \rangle - c)|\langle J \rangle_J|^2} \left[ \left\langle \frac{1}{|\lambda_{\text{isol}} - D|^2} \right\rangle_D^{-1} - c \left\langle |J|^2 \right\rangle_J \right]. \tag{S85}$$

Since $\langle r \rangle^2 \geq 0$, we obtain the following condition for the existence of the outlier:

$$\left\langle \frac{1}{|\lambda_{\text{isol}} - D|^2} \right\rangle_D^{-1} \geq c \left\langle |J|^2 \right\rangle_J. \tag{S86}$$

We compare Eq. (S86) with the Eq. (S69) for the support $\Omega$, and find that Eq. (S86) is equivalent to the condition $\lambda_{\text{isol}} \notin \Omega$. Hence, the outlier exists as long as it is located outside the absolute continuous part of the spectrum.

We find analytical expressions for the second moments $\langle r^2 \rangle$ and $\langle l^2 \rangle$ by substituting the expressions for $\langle r \rangle^2$ and $\langle l \rangle^2$, given by Eqs. (S84)-(S85), in Eqs. (S81) and (S82):

$$\langle r^2 \rangle = \frac{\langle J \rangle_J^2}{|\langle J \rangle_J|^2} \frac{\left[ \left\langle \frac{1}{|\lambda_{\text{isol}} - D|^2} \right\rangle_D^{-1} - c \left\langle |J|^2 \right\rangle_J \right]}{\left[ \left\langle \frac{1}{(\lambda_{\text{isol}} - D)^2} \right\rangle_D^{-1} - c \left\langle J^2 \right\rangle_J \right]}, \tag{S87}$$

$$\langle l^2 \rangle = \frac{\langle J^* \rangle_J^2}{|\langle J \rangle_J|^2} \frac{\left[ \left\langle \frac{1}{|\lambda_{\text{isol}} - D|^2} \right\rangle_D^{-1} - c \left\langle |J|^2 \right\rangle_J \right]}{\left[ \left\langle \frac{1}{(\lambda_{\text{isol}}^* - D^*)^2} \right\rangle_D^{-1} - c \left\langle J^2 \right\rangle_J^* \right]}. \tag{S88}$$

Since $\lambda_{\text{isol}}$ solves Eq. (S83), we find that $\lambda_{\text{isol}} \in \mathbb{R}$ as long as $J \in \mathbb{R}$ and $D \in \mathbb{R}$, which implies $\langle r^2 \rangle = \langle l^2 \rangle = 1$ from the above two equations. If $A_{jj} = 0$ for all values of $j$, and thus $p_D(D) = \delta(D)$, then $\lambda_{\text{isol}} = c \langle J \rangle_J$.

### B. Eigenvectors at the boundary of the support

The values $\lambda_{\text{b}}$, which determine the boundary of the support of the spectral density function, are given by the non-trivial solution to the Eqs. (S77)-(S82) for which the first moments are equal to zero, i.e., $\langle r \rangle = \langle l \rangle = 0$. From Eqs. (S79) and (S80) we find that $\lambda_{\text{b}}$ solves:

$$\left\langle \frac{1}{|\lambda_{\text{b}} - D|^2} \right\rangle_D = \frac{1}{c \langle |J|^2 \rangle_J}. \tag{S89}$$

Equation (S89) is also consistent with the Eq. (S69) for the support of the spectral density derived through a stability analysis. If diagonal elements are zero, i.e., $A_{jj} = 0$ for all values of $j$, then $|\lambda_{\text{b}}|^2 = c \langle |J|^2 \rangle_J$.

The eigenvector moments at the boundary of the support satisfy $\langle r^2 \rangle = \langle l^2 \rangle = 0$. Indeed, in this case Eqs. (S81) and (S82) become

$$\langle r^2 \rangle = c \langle r^2 \rangle \langle J^2 \rangle_J \left\langle \frac{1}{(\lambda_{\text{b}} - D)^2} \right\rangle_D, \tag{S90}$$

$$\langle l^2 \rangle = c \langle l^2 \rangle \langle J^2 \rangle_J^* \left\langle \frac{1}{(\lambda_b^* - D^*)^2} \right\rangle_D. \tag{S91}$$

Since $\lambda_{\text{b}}$ is in general a complex solution to Eq. (S89), we have that

$$c \langle J^2 \rangle_J \left\langle \frac{1}{(\lambda_{\text{b}} - D)^2} \right\rangle_D \neq 1, \tag{S92}$$

and therefore $\langle r^2 \rangle = \langle l^2 \rangle = 0$.

## S7. LAPLACIAN MATRICES ON AN ORIENTED GRAPH

In this subsection, we consider the solution to the Eqs. (S59)-(S64) for an ensemble of Laplacian random matrices defined on an oriented random graph. For Laplacian matrices the diagonal elements depend on the off-diagonal elements as follows:

$$J_{jj} = - \sum_{k \in \partial_j^{\text{out}}} J_{jk}. \tag{S93}$$

We thus have

$$p(D, J_1, J_2, \ldots, J_{K^{\text{out}}}, J'_1, J'_2, \ldots, J'_{K^{\text{in}}}) = \delta \left( D - \sum_{j=1}^{K^{\text{out}}} J_j \right) p(J_1, J_2, \ldots, J_{K^{\text{out}}}, J'_1, J'_2, \ldots, J'_{K^{\text{in}}}). \tag{S94}$$

For simplicity, let us further consider real off-diagonal elements that fulfill $J_{jk} \geq 0$. In this case, the Laplacian matrix represents the transition-rate matrix of a continuous-time Markov process and the constraint given by Eq. (S93) ensures the conservation of probability in the corresponding Master equation.

Equation (S93) breaks the symmetry of the ensemble with respect to its conjugate transpose: $P(\boldsymbol{A}_n) \neq P(\boldsymbol{A}_n^\dagger)$. This implies that $p_R(z) \neq p_L(z)$, and the right- and left-eigenvector distributions are different. For Laplacian matrices, the set of self-consistent equations in the eigenvector moments, given by Eqs. (S59)-(S64), are

$$\langle r \rangle = \langle r \rangle \left\langle \frac{\sum_{j=1}^{K^{\text{out}}} J_j}{\lambda + \sum_{j=1}^{K^{\text{out}}} J_j} \right\rangle, \tag{S95}$$

$$\langle l \rangle = \langle l \rangle \left\langle \frac{\sum_{j=1}^{K^{\text{in}}} J'_j}{\lambda^* + \sum_{j=1}^{K^{\text{out}}} J_j} \right\rangle, \tag{S96}$$

$$\langle |r|^2 \rangle = \langle |r|^2 \rangle \left\langle \frac{\sum_{j=1}^{K^{\text{out}}} J_j^2}{|\lambda + \sum_{j=1}^{K^{\text{out}}} J_j|^2} \right\rangle + |\langle r \rangle|^2 \left\langle \frac{\sum_{k \neq j=1}^{K^{\text{out}}} J_k J_j}{|\lambda + \sum_{j=1}^{K^{\text{out}}} J_j|^2} \right\rangle, \tag{S97}$$

$$\langle |l|^2 \rangle = \langle |l|^2 \rangle \left\langle \frac{\sum_{j=1}^{K^{\text{in}}} \left(J'_j\right)^2}{|\lambda + \sum_{j=1}^{K^{\text{out}}} J_j|^2} \right\rangle + |\langle l \rangle|^2 \left\langle \frac{\sum_{k \neq j=1}^{K^{\text{in}}} J'_k J'_j}{|\lambda + \sum_{j=1}^{K^{\text{out}}} J_j|^2} \right\rangle, \tag{S98}$$

$$\langle r^2 \rangle = \langle r^2 \rangle \left\langle \frac{\sum_{j=1}^{K^{\text{out}}} J_j^2}{(\lambda + \sum_{j=1}^{K^{\text{out}}} J_j)^2} \right\rangle + \langle r \rangle^2 \left\langle \frac{\sum_{k \neq j=1}^{K^{\text{out}}} J_k J_j}{(\lambda + \sum_{j=1}^{K^{\text{out}}} J_j)^2} \right\rangle, \tag{S99}$$

$$\langle l^2 \rangle = \langle l^2 \rangle \left\langle \frac{\sum_{j=1}^{K^{\text{in}}} \left(J'_j\right)^2}{(\lambda^* + \sum_{j=1}^{K^{\text{out}}} J_j)^2} \right\rangle + \langle l \rangle^2 \left\langle \frac{\sum_{k \neq j=1}^{K^{\text{in}}} J'_k J'_j}{(\lambda^* + \sum_{j=1}^{K^{\text{out}}} J_j)^2} \right\rangle. \tag{S100}$$

Here $\langle \ldots \rangle$ stands for the average over the joint distribution of $K^{\text{in}}$, $K^{\text{out}}$ and the off-diagonal matrix elements $\{J_j\}$ and $\{J'_j\}$. Note that here we have not normalized the eigenvector distribution (and we have thus not used Eq. (S76)). Just as for adjacency matrices, we find two non-trivial solutions to the Eqs. (S95)-(S100): one corresponding to the outlier $\lambda_{\text{isol}}$ and one to the boundary of the continuous part of the spectrum $\lambda_b$.

### A. The outlier eigenpair

According to the Perron-Frobenius theorem the outlier $\lambda_{\text{isol}} = 0$. This follows also from the Eq. (S95). Equation (S95) admits a nontrivial fixed-point solution, provided there is a value $\lambda = \lambda_{\text{isol}}$ such that

$$\left\langle \frac{\sum_{j=1}^{K^{\text{out}}} J_j}{\lambda_{\text{isol}} + \sum_{j=1}^{K^{\text{out}}} J_j} \right\rangle = 1. \tag{S101}$$

The simplest solution to the above equation is $\lambda_{\text{isol}} = 0$, which is indeed the outlier of a Laplacian matrix.

We use the normalization $\langle |r|^2 \rangle = 1$ and the value $\lambda_{\text{isol}} = 0$ in Eqs. (S97) and (S99), and we obtain $\langle r \rangle = \langle r^2 \rangle = 1$. These results are also consistent with the Perron-Frobenius theorem for Laplacian matrices, which implies that $p_R(r) = \delta(r; 1)$ [18].

We determine the moments of the left-eigenvector at the isolated eigenvalue $\lambda = \lambda_{\text{isol}} = 0$. We rewrite therefore the Eqs. (S96) and (S98) as follows

$$\langle l_t \rangle = \langle l_{t-1} \rangle \left\langle \frac{\sum_{j=1}^{K^{\text{in}}} J'_j}{\lambda^*_{\text{isol}} + \sum_{j=1}^{K^{\text{out}}} J_j} \right\rangle, \tag{S102}$$

$$\langle |l_t|^2 \rangle = \langle |l_{t-1}|^2 \rangle \left\langle \frac{\sum_{j=1}^{K^{\text{in}}} (J'_j)^2}{|\lambda_{\text{isol}} + \sum_{j=1}^{K^{\text{out}}} J_j|^2} \right\rangle + |\langle l_{t-1} \rangle|^2 \left\langle \frac{\sum_{k \neq j=1}^{K^{\text{in}}} J'_k J'_j}{|\lambda_{\text{isol}} + \sum_{j=1}^{K^{\text{out}}} J_j|^2} \right\rangle, \tag{S103}$$

with the discrete index $t$ denoting an iteration step. Note that since we can set $\langle |l_t|^2 \rangle = \langle l_t^2 \rangle$ and $\langle |l_t| \rangle = \langle l_t \rangle$, we use $\langle l_t^2 \rangle$ and $\langle l_t \rangle$ in the following. We use $\lambda_{\text{isol}} = 0$ to write:

$$\langle l_t \rangle = \langle l_{t-1} \rangle \left\langle \frac{\sum_{j=1}^{K^{\text{in}}} J'_j}{\sum_{j=1}^{K^{\text{out}}} J_j} \right\rangle, \tag{S104}$$

$$\langle l_t^2 \rangle = \langle l_{t-1}^2 \rangle \left\langle \frac{\sum_{j=1}^{K^{\text{in}}} (J'_j)^2}{|\sum_{j=1}^{K^{\text{out}}} J_j|^2} \right\rangle + \langle l_{t-1} \rangle^2 \left\langle \frac{\sum_{k \neq j=1}^{K^{\text{in}}} J'_k J'_j}{|\sum_{j=1}^{K^{\text{out}}} J_j|^2} \right\rangle. \tag{S105}$$

We characterize the asymptotic fixed-point behaviour of the moments by normalizing $\langle l_t \rangle$ as follows:

$$\frac{\langle l_t^2 \rangle}{\langle l_t \rangle^2} = \frac{\langle l_{t-1}^2 \rangle}{\langle l_t \rangle^2} \left\langle \frac{\sum_{j=1}^{K^{\text{in}}} (J'_j)^2}{(\sum_{j=1}^{K^{\text{out}}} J_j)^2} \right\rangle + \frac{\langle l_{t-1} \rangle^2}{\langle l_t \rangle^2} \left\langle \frac{\sum_{k \neq j=1}^{K^{\text{in}}} J'_k J'_j}{(\sum_{j=1}^{K^{\text{out}}} J_j)^2} \right\rangle. \tag{S106}$$

We substitute Eq. (S104) in the above and we find the recursion

$$\frac{\langle l_t^2 \rangle}{\langle l_t \rangle^2} = \frac{\langle l_{t-1}^2 \rangle}{\langle l_{t-1} \rangle^2} \left\langle \frac{\sum_{j=1}^{K^{\text{in}}} J'_j}{\sum_{j=1}^{K^{\text{out}}} J_j} \right\rangle^{-2} \left\langle \frac{\sum_{j=1}^{K^{\text{in}}} (J'_j)^2}{(\sum_{j=1}^{K^{\text{out}}} J_j)^2} \right\rangle + \left\langle \frac{\sum_{j=1}^{K^{\text{in}}} J'_j}{\sum_{j=1}^{K^{\text{out}}} J_j} \right\rangle^{-2} \left\langle \frac{\sum_{k \neq j=1}^{K^{\text{in}}} J'_k J'_j}{(\sum_{j=1}^{K^{\text{out}}} J_j)^2} \right\rangle. \tag{S107}$$

The ratio $\frac{\langle l^2(t) \rangle}{\langle l(t) \rangle^2}$ converges to the value:

$$\frac{\langle l^2 \rangle}{\langle l \rangle^2} = \frac{\left\langle \frac{\sum_{k \neq j=1}^{K^{\text{in}}} J'_k J'_j}{\left(\sum_{j=1}^{K^{\text{out}}} J_j\right)^2} \right\rangle}{\left\langle \frac{\sum_{j=1}^{K^{\text{in}}} J'_j}{\sum_{j=1}^{K^{\text{out}}} J_j} \right\rangle^2 - \left\langle \frac{\sum_{j=1}^{K^{\text{in}}} (J'_j)^2}{\left(\sum_{j=1}^{K^{\text{out}}} J_j\right)^2} \right\rangle}, \tag{S108}$$

for $t \to \infty$, and for values $\left\langle \frac{\sum_{j=1}^{K^{\text{in}}} J'_j}{\sum_{j=1}^{K^{\text{out}}} J_j} \right\rangle^2 > \left\langle \frac{\sum_{j=1}^{K^{\text{in}}} (J'_j)^2}{\left(\sum_{j=1}^{K^{\text{out}}} J_j\right)^2} \right\rangle$. If $\left\langle \frac{\sum_{j=1}^{K^{\text{in}}} J'_j}{\sum_{j=1}^{K^{\text{out}}} J_j} \right\rangle^2 \leq \left\langle \frac{\sum_{j=1}^{K^{\text{in}}} (J'_j)^2}{\left(\sum_{j=1}^{K^{\text{out}}} J_j\right)^2} \right\rangle$, then $\langle l \rangle = 0$ and $\langle l^2 \rangle \neq 0$.

For the particular case for which all off-diagonal elements are equal to one, i.e., $J_j = 1$, as considered in the main text, we have:

$$\frac{\langle l^2 \rangle}{\langle l \rangle^2} = \frac{\left\langle \frac{(K^{\text{in}})^2 - c}{(K^{\text{out}})^2} \right\rangle}{\left\langle \frac{c}{K^{\text{out}}} \right\rangle^2 - \left\langle \frac{c}{(K^{\text{out}})^2} \right\rangle}. \tag{S109}$$

### B. Eigenvectors at the boundary of the support

Tha values $\lambda_b$ for the boundary of the support of the spectral density function, $\lambda_b \in \partial\Omega$, are given by the non-trivial solution to the Eqs. (S97)-(S100) for which $\langle r \rangle = \langle l \rangle = 0$.

The existence of a non-trivial solution depends now on the second moments $\langle |r|^2 \rangle$ and $\langle |l|^2 \rangle$. Substituting $\langle r \rangle = \langle l \rangle = 0$ in Eqs. (S97)-(S100), we obtain a nontrivial second moment $\langle |r|^2 \rangle \neq 0$ for the values $\lambda = \lambda_b^{(r)}$ where the

following equation holds

$$\left\langle \frac{\sum_{j=1}^{K^{\text{out}}} J_j^2}{|\lambda_b^{(r)} + \sum_{j=1}^{K^{\text{out}}} J_j|^2} \right\rangle = 1. \tag{S110}$$

In addition, a nontrivial second moment $\langle |l|^2 \rangle \neq 0$ is obtained for the values $\lambda = \lambda_b^{(l)}$ at which the following equation is fulfilled

$$\left\langle \frac{\sum_{j=1}^{K^{\text{in}}} \left(J_j'\right)^2}{|\lambda_b^{(l)} + \sum_{j=1}^{K^{\text{out}}} J_j|^2} \right\rangle = 1. \tag{S111}$$

The Eqs. (S110)-(S111) are consistent with the equations (S65) and (S66) which we have derived from a local stability analysis of the resolvent equations.

## S8. REGULAR NON-ORIENTED RANDOM MATRIX ENSEMBLE

Out theory, based on Eqs.(5)-(10), holds for local-tree like matrix ensembles, and is not restricted to oriented ensembles. We illustrate here the application of our theory to a non-oriented random matrix ensemble. We derive an exact analytical expression for the eigenvalue outlier.

We consider here a $2c$-regular ensemble of non-Hermitian matrices for which each vertex $j$ has exactly $2c$ non-zero off-diagonal elements $A_{jk}$; for each vertex $j$, half of the non-zero offdiagonal elements are of the form $(A_{jk}, A_{kj}) = (p_+ e^{i\phi_{kj}}, p_- e^{-i\phi_{kj}})$, and the other half is of the form $(A_{jk}, A_{kj}) = (p_- e^{-i\phi_{jk}}, p_+ e^{i\phi_{jk}})$, with $p_+, p_- \geq 0$. The phase variables $\phi_{kj}$ are random, independent and identically distributed with distribution $p(\phi)$. We call this ensemble the *elliptic $2c$-regular ensemble*, or more briefly, the *elliptic regular ensemble*.

For $p_- = 0$ the elliptic regular ensemble is an oriented matrix ensemble, similar to the ones discussed above in Section S6, whereas for $p_+ = p_-$ and $p(\phi) = \delta(\phi - \pi)(1 - \Delta)/2 + \delta(\phi)(1 + \Delta)/2$, with $\Delta \in [-1, 1]$, the elliptic regular ensemble is identical to the real and symmetric ensemble studied in [9, 19]. We define the asymmetry parameter $\epsilon = p_-/p_+$. For a phase distribution $p(\phi) = \delta(\phi - \pi)(1 - \Delta)/2 + \delta(\phi)(1 + \Delta)/2$, the elliptic regular ensemble interpolates between an oriented regular ensemble ($\epsilon = 0$) and a symmetric regular ensemble ($\epsilon = 1$)

### A. Spectral density

An analytical expression for the spectral density of the elliptic regular ensemble has been determined in [4]. For large connectivities $c \gg 1$ the spectral density is given by the elliptical law [20], which clarifies why we call the ensemble the elliptic regular ensemble. For finite $c$, the support $\Omega$ is given by the set of values $\lambda = x + iy$ for which [4]

$$\left(\frac{x}{a}\right)^2 + \left(\frac{y}{b}\right)^2 < 1 \quad, \tag{S112}$$

and with parameters

$$a = \sqrt{H} + (2c - 1)\frac{p_+ p_-}{\sqrt{H}} \quad, \tag{S113}$$

$$b = \sqrt{H} - (2c - 1)\frac{p_+ p_-}{\sqrt{H}} \quad, \tag{S114}$$

$$2H = c\left(p_+^2 + p_-^2\right) + \sqrt{c^2\left(p_+^2 - p_-^2\right)^2 + 4(c - 1)^2 (p_+ p_-)^2} \quad. \tag{S115}$$

Notice that the support is independent of the distribution $p(\phi)$ of phase variables.

### B. Outliers of the ensemble

The novelty of our work is that we can compute outliers of sparse non-Hermitian ensembles. To this aim, we apply the Eqs. (5)-(10) to elliptic regular ensemble. For the elliptic regular ensemble, the resolvent Eqs. (8) read

$$g^{(+)} = -\left(\lambda + c\, p_+ p_- g^{(+)} + (c - 1)\, p_+ p_- g^{(-)}\right)^{-1} \quad, \tag{S116}$$

$$g^{(-)} = -\left(\lambda + (c - 1)\, p_+ p_- g^{(+)} + c\, p_+ p_- g^{(-)}\right)^{-1} \quad, \tag{S117}$$

and the Eqs. (7) read

$$g = -\left(\lambda + c\, p_+ p_- g^{(+)} + c\, p_+ p_- g^{(-)}\right)^{-1} \quad . \tag{S118}$$

The solution to the Eqs. (S116)-(S118) is

$$g^{(+)} = g^{(-)} = \tilde{g} \quad , \tag{S119}$$

$$\tilde{g} = \frac{-\lambda + \sqrt{\lambda^2 - 4(2c-1)p_- p_+}}{2(2c-1)p_- p_+} \quad , \tag{S120}$$

$$g = -\frac{2c-1}{\lambda(c-1) + c\sqrt{\lambda^2 - 4(2c-1)p_- p_+}} \quad . \tag{S121}$$

The random variables $r_j^{(\ell)}$ in Eq. (9) solve

$$r_j^{(\ell)} = -\tilde{g} \sum_{k \in \partial_j \setminus \{\ell\}} A_{jk} r_k^{(j)}, \tag{S122}$$

and a similar equation holds for the random variables $l_j^{(\ell)}$. The right eigenvector elements $r_j$ solve:

$$r_j = -g \sum_{k \in \partial_j} A_{jk} r_k^{(j)} . \tag{S123}$$

We take the ensemble average of the Eqs.(S122) to find

$$\langle r \rangle_+ = -\tilde{g}\, p_+ c \langle e^{i\phi} \rangle \langle r \rangle_+ - \tilde{g}\, p_-(c-1) \langle e^{-i\phi} \rangle \langle r \rangle_- \quad , \tag{S124}$$

$$\langle r \rangle_- = -\tilde{g}\, p_+(c-1) \langle e^{i\phi} \rangle \langle r \rangle_+ - \tilde{g}\, p_- c \langle e^{-i\phi} \rangle \langle r \rangle_- \quad , \tag{S125}$$

and

$$\langle r \rangle = -g\, p_+ c \langle e^{i\phi} \rangle \langle r \rangle_+ - g\, p_- c \langle e^{-i\phi} \rangle \langle r \rangle_- \quad . \tag{S126}$$

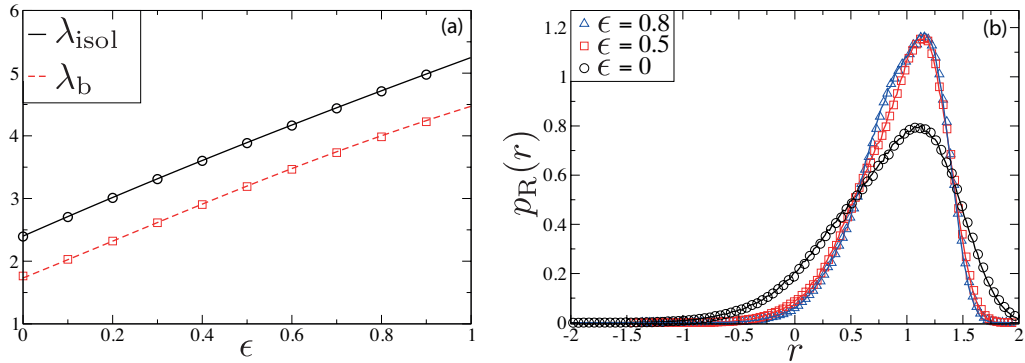

FIG. S1: Eigenvalue outlier $\lambda_{\rm isol}$, the boundary $\lambda_{\rm b}$ of the support $\Omega$ on the real axis, and the distribution $p_{\rm R}$ of right-eigenvector elements associated to the outlier, for the elliptic regular ensemble with a phase distribution $p(\phi) = (1-\Delta)/2\,\delta(\phi-\pi) + (1+\Delta)/2\,\delta(\phi)$, and with parameters $\Delta = 0.8$, $p_+ = 1$, $p_- = \epsilon$ and $c = 3$. Subfigure (a): we compare direct-diagonalization results of $\lambda_{\rm isol}$ and $\lambda_{\rm b}$ for finite-sized matrices ($n = 1000$, markers) with the theoretical results for infinite-sized matrices, given by Eqs. (S129)-(S130) (lines). Direct-diagonalization results are averages over 1000 samples. Subfigure (b): distribution $p_{\rm R}$ of right-eigenvector elements associated to the outlier $\lambda_{\rm isol}$ for three values of the asymmetry parameter $\epsilon$. We compare direct-diagonalization results of 1000 samples of finite-sized matrices ($n = 1000$, makers) with theoretical results of infinite-sized matrices obtained from solving the Eqs. (S122)-(S123) with the population dynamics algorithm (solid lines). The distributions are normalized such that $\langle r^2 \rangle = 1$ and ${\rm Im}\,(\langle r \rangle) = 0$.

The outliers $\lambda_{\rm isol}$ are given by the values $\lambda$ for which the Eqs. (S124)-(S125) provide a non-trivial solution. After some algebra we find

$$\lambda_{\rm isol} = \frac{\tilde{\lambda}^2 + (2c-1)\epsilon}{\tilde{\lambda}} \quad , \tag{S127}$$

with

$$\tilde{\lambda} = \frac{1}{2}\left(p_+ c \langle e^{i\phi}\rangle + p_- c \langle e^{-i\phi}\rangle + \sqrt{(p_+ c \langle e^{i\phi}\rangle - p_- c \langle e^{-i\phi}\rangle)^2 + 4p_+ p_-(c-1)^2 \langle e^{i\phi}\rangle \langle e^{-i\phi}\rangle}\right) \quad . \tag{S128}$$

The outlier $\lambda_{\rm isol}$ exists as long as $\lambda_{\rm isol} \notin \Omega$. Let us consider a particular example. We set $p_+ = 1$ and $p_- = \epsilon$, with $\epsilon \in [0,1]$, and consider a distribution $p(\phi) = (1-\Delta)/2\,\delta(\phi-\pi) + (1+\Delta)/2\,\delta(\phi)$, with $\Delta \in [-1,1]$. We then find for the outlier $\lambda_{\rm isol}$ the expression:

$$\lambda_{\rm isol} = \frac{\Delta}{2}\left(c(1+\epsilon) + \sqrt{c^2(1-\epsilon)^2 + 4\epsilon(c-1)^2}\right) + \frac{2(2c-1)\epsilon}{\Delta\left(c(1+\epsilon) + \sqrt{c^2(1-\epsilon)^2 + 4\epsilon(c-1)^2}\right)} \quad , \tag{S129}$$

which exists as long as $\lambda_{\rm isol} \notin \Omega$. Note that Eq. (S129) interpolates between the value $\lambda_{\rm isol} = c\Delta$ for oriented ensembles ($\epsilon = 0$, see Eq. (14)), and the value $\lambda_{\rm isol} = \Delta(2c-1) + \frac{1}{\Delta}$ for symmetric ensembles ($\epsilon = 1$, see [9, 19]). Using Eqs. (S112)-(S115) we find the condition $\lambda_{\rm isol} \geq \lambda_b$, with

$$\lambda_b = \sqrt{H} + (2c-1)\epsilon/\sqrt{H} \tag{S130}$$

the boundary of the support on the real axis, and $2H = c\left(1+\epsilon^2\right) + \sqrt{c^2\left(1-\epsilon^2\right)^2 + 4(c-1)^2\epsilon^2}$.

We compare in figure S1(a) the analytical expressions for $\lambda_{\rm isol}$ and $\lambda_b$, given by Eqs. (S129)-(S130), with direct diagonalization results of finite-sized matrices; we plot $\lambda_{\rm isol}$ and $\lambda_b$ as a function of the asymmetry parameter $\epsilon$. Our analytical expressions are in perfect agreement with direct diagonalization results. Finally, in Fig. S1(b) we compare the numerical solution of Eqs. (S122) and (S123) using the population dynamics algorithm with direct diagonalization results; theoretical results are in perfect agreement with direct diagonalization results.

---

[1] J. Feinberg and A. Zee, "Non-gaussian non-hermitian random matrix theory: phase transition and addition formalism," *Nuclear Physics B*, vol. 501, no. 3, pp. 643–669, 1997.
[2] J. Feinberg and A. Zee, "Non-hermitian random matrix theory: method of hermitian reduction," *Nuclear Physics B*, vol. 504, no. 3, pp. 579–608, 1997.
[3] T. Rogers and I. P. Castillo, "Cavity approach to the spectral density of non-hermitian sparse matrices," *Phys. Rev. E*, vol. 79, p. 012101, Jan 2009.
[4] I. Neri and F. L. Metz, "Spectra of sparse non-hermitian random matrices: An analytical solution," *Phys. Rev. Lett.*, vol. 109, p. 030602, Jul 2012.
[5] F. L. Metz, I. Neri, and D. Bollé, "Spectra of sparse regular graphs with loops," *Phys. Rev. E*, vol. 84, p. 055101, Nov 2011.
[6] D. Bollé, F. L. Metz, and I. Neri, *Spectral Analysis, Differential Equations and Mathematical Physics: A Festschrift in Honor of Fritz Gesztesy's 60th Birthday*, ch. On the spectra of large sparse graphs with cycles. 2013.
[7] Y. Weiss and W. T. Freeman, "Correctness of belief propagation in gaussian graphical models of arbitrary topology," *Neural computation*, vol. 13, no. 10, pp. 2173–2200, 2001.
[8] D. Bickson, "Gaussian belief propagation: Theory and aplication," *arXiv preprint arXiv:0811.2518*, 2008.
[9] Y. Kabashima, H. Takahashi, and O. Watanabe, "Cavity approach to the first eigenvalue problem in a family of symmetric random sparse matrices," *Journal of Physics: Conference Series*, vol. 233, no. 1, p. 012001, 2010.
[10] T. Rogers, I. P. Castillo, R. Kühn, and K. Takeda, "Cavity approach to the spectral density of sparse symmetric random matrices," *Phys. Rev. E*, vol. 78, p. 031116, Sep 2008.
[11] J. W. Negele and H. Orland, *Quantum many-particle systems*, vol. 200. Addison-Wesley New York, 1988.
[12] R. Abou-Chacra, D. J. Thouless, and P. W. Anderson, "A selfconsistent theory of localization," *Journal of Physics C: Solid State Physics*, vol. 6, no. 10, p. 1734, 1973.
[13] F. L. Metz, I. Neri, and D. Bollé, "Localization transition in symmetric random matrices," *Physical Review E*, vol. 82, no. 3, p. 031135, 2010.
[14] G. Biroli, G. Semerjian, and M. Tarzia, "Anderson model on bethe lattices: density of states, localization properties and isolated eigenvalue," *Progress of Theoretical Physics Supplement*, vol. 184, pp. 187–199, 2010.
[15] I. Neri and D. Bollé, "The cavity approach to parallel dynamics of ising spins on a graph," *Journal of Statistical Mechanics: Theory and Experiment*, vol. 2009, no. 08, p. P08009, 2009.



\end{document}